\documentclass{article}

\usepackage{microtype}
\usepackage{graphicx}
\usepackage{subfigure}
\usepackage{booktabs} 

\usepackage{hyperref}

\usepackage[accepted]{icml2025}

\usepackage{amsmath}
\usepackage{amssymb}
\usepackage{mathtools}
\usepackage{amsthm}

\usepackage[capitalize,noabbrev]{cleveref}

\usepackage{hyperref}
\usepackage{url}
\usepackage{amssymb}
\usepackage{booktabs}
\usepackage{multirow}
\usepackage{pifont}
\usepackage{colortbl}
\usepackage{wrapfig}
\usepackage{graphicx}
\usepackage[breakable]{tcolorbox}

\definecolor{darkgreen}{rgb}{0,0.8,0}
\newcommand{\pos}[1]{{\tiny\textcolor{green!60!black}{\,\,$+#1$}}}
\newcommand{\negd}[1]{{\tiny\textcolor{red!70!black}{\,\,$-#1$}}}
\newcommand{\posrev}[1]{{\tiny\textcolor{red!70!black}{\,\,$+#1$}}}
\newcommand{\negdrev}[1]{{\tiny\textcolor{green!60!black}{\,\,$-#1$}}}
\newcommand{\zerod}{{\tiny\phantom{+}\textcolor{gray}{\,\,\,$0.00$}}}

\theoremstyle{plain}

\theoremstyle{definition}

\theoremstyle{remark}

\usepackage[textsize=tiny]{todonotes}
\usepackage{xspace}

\newcommand{\methodname}{SWE-Playground\xspace}

\usepackage{color-edits}
\addauthor{gn}{magenta}
\addauthor{yiqi}{blue}
\addauthor{ag}{orange}

\icmltitlerunning{Training Versatile Coding Agents in Synthetic Environments}

\begin{document}

\twocolumn[
\icmltitle{Training Versatile Coding Agents in Synthetic Environments}



\icmlsetsymbol{equal}{*}

\begin{icmlauthorlist}
\icmlauthor{Yiqi Zhu}{equal,thu}
\icmlauthor{Apurva Gandhi}{cmu}
\icmlauthor{Graham Neubig}{cmu}
\end{icmlauthorlist}

\icmlaffiliation{thu}{Tsinghua University}
\icmlaffiliation{cmu}{Carnegie Mellon University}

\icmlcorrespondingauthor{Graham Neubig}{gneubig@cs.cmu.edu}

\icmlkeywords{Machine Learning, ICML}

\vskip 0.3in
]



\printAffiliationsAndNotice{\icmlEqualContribution} 

\begin{abstract}

Prior works on training software engineering agents have explored utilizing existing resources such as issues on GitHub repositories to construct software engineering tasks and corresponding test suites.
These approaches face two key limitations: (1) their reliance on pre-existing GitHub repositories offers limited flexibility, and (2) their primary focus on issue resolution tasks restricts their applicability to the much wider variety of tasks a software engineer must handle.
To overcome these challenges, we introduce \methodname, a novel pipeline for generating environments and trajectories which supports the training of versatile coding agents. Unlike prior efforts, \methodname synthetically generates projects and tasks from scratch with strong language models and agents, eliminating reliance on external data sources. This allows us to tackle a much wider variety of coding tasks, such as reproducing issues by generating unit tests and implementing libraries from scratch. We demonstrate the effectiveness of this approach on three distinct benchmarks, and results indicate that \methodname produces trajectories with dense training signal, enabling agents to reach comparable performance with significantly fewer trajectories than previous works. \texttt{Project Page: \href{https://neulab.github.io/SWE-Playground}{neulab.github.io/SWE-Playground}}

\end{abstract}

\begin{figure*}[h!]
    \centering
    \includegraphics[width=0.9\textwidth]{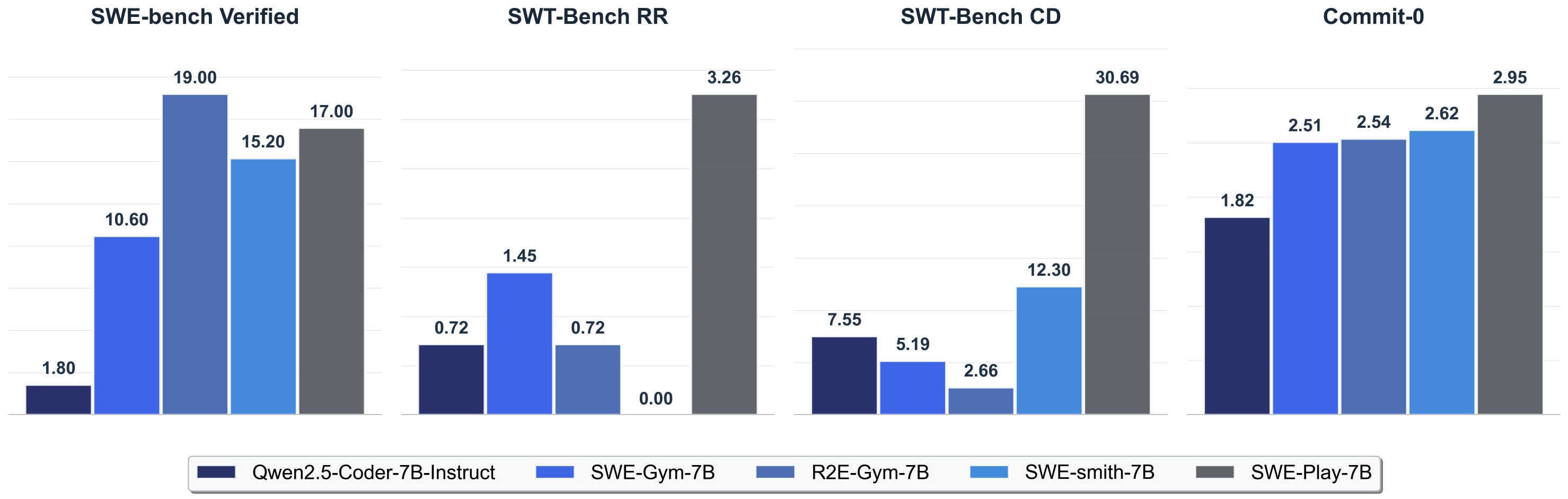}
    \vspace{-0.5em}
    \caption{Overview of our \methodname against previous methods on diverse coding benchmarks. As the results indicate, though our model falls slightly short on SWE-bench Verified (using fewer training trajectories against R2E-Gym and SWE-smith), it outperforms all baselines on the other benchmarks and metrics, demonstrating that our method is capable of training versatile coding agents.}
    \label{fig:overview}
    \vspace{-0.5em}
\end{figure*}

\section{Introduction}

Recent advances in Large Language Models (LLMs) have empowered Language Model Agents (LM Agents) to address a wide array of real-world tasks, with coding being a prime example~\citep{jimenez2024swebench, zhuo2024bigcodebench, jain2024livecodebench}. In an effort to enhance the coding proficiency of LLMs and agents, prior studies~\citep{pan2025training, jain2025r2e, yang2025swesmith} have explored creating specialized training environments designed to develop state-of-the-art open-weight software engineering (SWE) agents. These approaches utilize automated pipelines to collect SWE tasks with verifiable reward signals from existing Github repositories, enabling the training of SWE agents through agentic distillation or reinforcement learning (RL)~\citep{deepswe2025}.

However, as \citet{pan2025training} has demonstrated, the performance of SWE agents is highly dependent on the scale of the training data. Though pipelines of \citet{jain2025r2e} and \citet{yang2025swesmith} mitigate the reliance on human-written pull requests, \textbf{they remain fundamentally constrained by their reliance on existing GitHub repositories, thereby hindering the potential of creating larger-scale training datasets while maintaining high quality and flexibility}. Furthermore, real-world coding spans a diverse spectrum of tasks, including resolving issues~\citep{jimenez2024swebench}, generating unit tests to detect or reproduce potential problems~\citep{mundler2024swtbench}, and developing new projects from scratch~\citep{zhao2024commit0librarygenerationscratch}. An ideal coding agent should be capable of handling these varied tasks, yet \textbf{all aforementioned works predominantly focus on issue-resolution tasks while neglecting other critical scenarios}. As illustrated in Figure~\ref{fig:overview}, agents trained in these environments exhibit less pronounced or even deteriorated performance on other well-established benchmarks, despite their impressive results on SWE-bench.

In this paper, we introduce \methodname, a fully automated pipeline that synthetically constructs software engineering environments, including training tasks, starter code, and unit tests, from the scratch, for developing versatile coding agents. Using frontier LLMs and agents, the pipeline proposes a full software project, decomposes it into verifiable tasks, and automatically creates the repository scaffold along with matching unit tests. Implementations are then evaluated against these tests, providing a reliable reward signal for training agents.

\textbf{\methodname offers flexibility and extensibility.} A key advantage of \methodname lies in its inherent flexible framework. By synthetically creating all tasks and verification tests from scratch, our pipeline avoids any reliance on external resources. To demonstrate this adaptability, we extend our pipeline to support issue resolution and reproduction tasks via issue proposal and injection, and to create library generation tasks by stubbing out the generated implementation. This approach enables the collection of a rich and diverse training dataset, containing trajectories for both de novo project development and targeted issue resolution and reproduction. Since the entire pipeline operates without manual intervention and can be readily adapted to new tasks, often requiring little more than writing a different prompt to target a new domain, it presents a more scalable and cost-effective paradigm for wider adaption in the training of coding agents.

\textbf{\methodname supports training versatile coding agents.} Using trajectories collected exclusively from \methodname, we train models to validate the effectiveness of our approach. Our model consistently demonstrates superior performance over the base model across three distinct benchmarks that assess a wide range of coding capabilities. In stark contrast, agents trained on prior environments, such as SWE-Gym~\citep{pan2025training} and R2E-Gym~\citep{jain2025r2e}, which are typically limited to SWE-bench format tasks, exhibit limited generalization to these benchmarks. While they show minor gains on some out-of-domain tasks, these improvements do not match their source domain performance, and they suffer noticeable degradation on others. This result highlights that though prior environments successfully produce specialized agents that excels at SWE-bench, our \methodname is capable of cultivating robust and generalizable coding proficiency, and thus developing versatile coding agents.

\textbf{\methodname produces trajectories with dense training signal.} Our experiments collect 704 trajectories, a set significantly smaller than those used by R2E-Gym~\citep{jain2025r2e} and SWE-smith~\citep{yang2025swesmith}. Despite this smaller dataset, our models achieve comparable performance. Further analysis reveals that our trajectories are substantially richer, containing more tokens and tool calls on average compared to those from previous environments. Moreover, they feature a higher proportion of bash execution, demonstrating a paradigm of execution-based software development rather than simple code generation. We attribute this high data efficiency to the dense and complex nature of \methodname tasks and generated trajectories.

\section{What Makes a Good Software Agent?}

\subsection{Necessary Skills and Evaluation Benchmarks}
\label{sec:codingskills}

Software engineering is a multifaceted discipline that extends far beyond merely writing code, encompassing tasks such as designing, coding, testing, and debugging. Underscoring this, \citet{meyer2019today} studies the work of human software developers and finds that only 15\% of their time is spent actively coding. Moreover, code development itself is not monolithic, and it requires a diverse set of specialized skills for effective software engineering.

Fortunately, there are a variety of benchmarks designed to evaluate the capabilities of language models to tackle these various coding tasks.
SWE-bench~\citep{jimenez2024swebench} is the canonical benchmark for the ``issue in, patch out'' bug-fixing use case, focusing on popular Python repositories.
SWE-bench-Live~\citep{zhang2025swebenchgoeslive} addresses the static limitations of SWE-bench, introducing a live and automated pipeline that curates fresh, contamination-resistant tasks from real-world GitHub issues.
SWE-Bench Pro~\citep{deng2025swebenchproaiagents} also builds upon SWE-bench with more challenging tasks to capture realistic, complex, enterprise-level problems beyond the scope of SWE-bench.

In addition to this specific issue resolution paradigm, benchmarks are emerging to cover other critical aspects of the code development lifecycle:
\begin{itemize}
    \vspace{-1em}
    \item \textbf{Building Libraries from Scratch:} Commit-0~\citep{zhao2024commit0librarygenerationscratch} tasks models with rebuilding and generating the entire library from the initial commit.
    \vspace{-0.5em}
    \item \textbf{Multilingual Bug Fixing:} Multi-SWE-bench ~\citep{zan2025multiswebench} and SWE-bench Multilingual~\citep{yang2025swesmith} cover a wider range programming languages beyond Python, including C++, TypeScript, Rust, etc.
    \vspace{-0.5em}
    \item \textbf{Multimodal Understanding:} ArtifactsBench~\citep{zhang2025artifactsbenchbridgingvisualinteractivegap} and SWE-bench Multimodal~\citep{yang2024swebenchmultimodal} test the capability to ground code generation and understanding in visual artifacts.
    \vspace{-0.5em}
    \item \textbf{Test Generation:} SWT-Bench~\citep{mundler2024swtbench} and TestGenEval~\citep{jain2025testgenevalrealworldunit} evaluates model performance on reproducing GitHub issues in Python. 
    \vspace{-0.5em}
    \item \textbf{Performance Optimization:} SWE-Perf~\citep{he2025sweperf}, SWE-fficiency~\citep{ma2025swefficiencylanguagemodelsoptimize}, and KernelBench~\citep{ouyang2025kernelbenchllmswriteefficient} benchmark models on optimizing code performance in diverse scenarios and programming languages.
\end{itemize}

\vspace{-1em}
A full comprehensive review can be found in Appendix~\ref{sec:existingbenchmarks}.

\subsection{Training Environments}

Several recent works have focused on developing specialized environments for training SWE agents. SWE-Gym~\citep{pan2025training} first explores the construction of environments for training SWE agents with pre-installed dependencies and executable test verification derived from real-world GitHub issues. R2E-Gym~\citep{jain2025r2e} proposes a synthetic data curation recipe to create execution environments from commits through backtranslation and test collection or generation, which removes the reliance on human-written pull requests or unit tests. Similarly, SWE-smith~\citep{yang2025swesmith} facilitates automatic generation of scalable training environments and instances via function rewriting and bug combination. SWE-Mirror~\citep{wang2025swemirrorscalingissueresolvingdatasets} further scales this by mirroring issues across repositories. SWE-Flow~\citep{zhang2025sweflow} utilizes Test-Driven Development (TDD) to produce structured development tasks with explicit goals directly from unit tests, thereby generating training instances. 

However, crucially, while these environments represent significant progress, they share a fundamental limitation that \textbf{they largely rely on existing GitHub repositories to source their data}. This approach inherently restricts the diversity and availability of tasks to what can be mined from public code. Consequently, it remains difficult to acquire balanced training data that covers the full spectrum of aforementioned skills for training a versatile coding agent. This bottleneck can be exacerbated by the extensive manual effort required for filtering, verifying, and packaging these real-world issues into training environments. Leveraging a fully automated pipeline that spans from project proposal to functionality implementation, \methodname attempts to address this fundamental problem.

\section{\methodname}

\begin{figure*}[h!]
    \centering
    \includegraphics[width=1.0\textwidth]{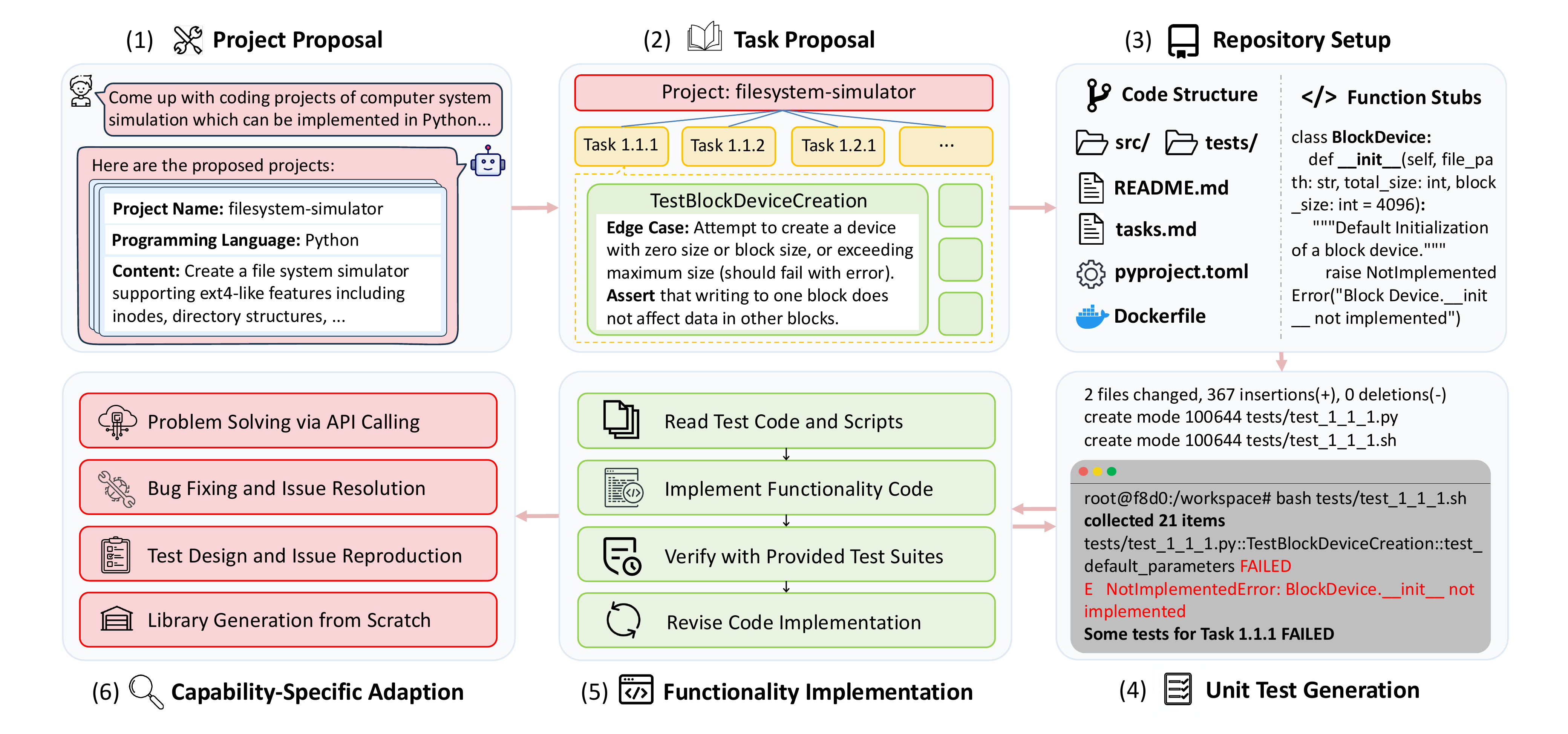}
    \vspace{-2em}
    \caption{Overview of the \methodname data generation pipeline. SWE-Playground (1) proposes a project that adheres to various requirements as specified in the prompt (Section~\ref{subsection:taskproposal}); (2) decomposes the project into implementable phases, modules and tasks along with checklists for necessary tests (Section~\ref{subsection:projectproposal}); (3) establishes repository structure to scaffold the implementation (Section~\ref{subsection:repositorysetup}); (4) generates unit tests with OpenHands agent based on the derived documentation (Section~\ref{subsection:unittestgeneration}); (5) utilizes generated unit tests as reward signal to collect functionality implementation trajectories (Section~\ref{subsection:functionalityimplementation}). Finally our pipeline can be (6) adapted for generating tasks for training models with diverse coding capabilities (Section~\ref{subsection:taskspecificgeneration}).}
    \label{fig:pipeine}
    \vspace{-0.5em}
\end{figure*}

In this section, we elaborate on the pipeline of \methodname and how we can collect high-quality data for training versatile coding agents with our proposed pipeline. The pipeline is illustrated by Figure~\ref{fig:pipeine}, and prompts used can be found in \autoref{sec:prompt}.

\subsection{Project Proposal}
\label{subsection:projectproposal}

The pipeline of \methodname begins with an LLM call requesting a project proposal aligned with a specific domain of coding problems we aim to generate for. This generation process is steered by a prompt that rigorously defines various parameters of the project. In our implementation, as detailed in Appendix~\ref{subsection:promptprojectproposal}, we prompt the agent to achieve particular high-level attributes such as project topic and programming language, and granular, robustness-ensuring constraints that are critical for maintaining quality, including:

\begin{itemize}
    \vspace{-1em}
    \item \textbf{Scope and complexity metrics} to guarantee scale, specifically requiring projects to involve multi-component architectures and a minimum volume of core logic rather than simple.
    \vspace{-0.5em}
    \item \textbf{Constraints on dependencies} that explicitly forbid high-level libraries for core tasks, forcing the agent to implement algorithms and data structures from scratch to demonstrate deep understanding.
    \vspace{-0.5em}
    \item \textbf{CLI-based interaction} to mandate a text-based, scriptable interface without GUI components, ensuring that all agent outputs are amenable to automated testing.
    \vspace{-0.5em}
    \item \textbf{Unambiguous specification protocols} requiring concrete IO formats and clear success criteria to prevent open-ended or subjective problem definitions.
    \vspace{-0.5em}
    \item \textbf{High algorithmic density} to filter out basic CRUD applications in favor of tasks requiring sophisticated logic, such as custom parsers, optimization engines, or compiler implementations.
\end{itemize}

\vspace{-0.5em}
Crucially, these constraints are highly adaptable. By simply modifying the system prompt, the requirements can be tailored to suit a wide range of use cases, ensuring the flexibility of our pipeline. Furthermore, to prevent mode collapse and ensure broad coverage, we require the model to generate multiple tasks simultaneously, thereby maximizing the diversity of the resulting project pool.

\subsection{Task Proposal}
\label{subsection:taskproposal}

Following the proposal stage, each project is processed independently to decompose the high-level requirements into executable units. To ensure a logical and organized implementation flow, we employ a hierarchical generation structure: the project is first divided into distinct phases, then modules, and finally concrete tasks, simulating the real-world project implementation process.

In this stage, achieving optimal task granularity is critical. We instruct the model to maintain a balanced workload by defining tasks at the granularity of a single pull request, ensuring they are substantial yet manageable. For example, typical tasks include scaffolding base data structures, implementing core algorithms, or developing user interface components. Furthermore, to maintain a focus on high-value engineering, we restrict the hierarchy to a maximum of five phases and explicitly exclude auxiliary activities such as documentation or deployment, forcing the agent to concentrate solely on building core functionality with clear logical dependency chains.

Once a task is defined, we generate a corresponding detailed checklist that provides a granular explanation and explicitly specifies the requisite unit tests, standard cases, specific assertions, and potential edge cases. This practice offers two primary advantages. First, unit tests serve as a reliable reward signal. Though previous works have explored training software engineering verifier models for inference-time scaling~\citep{pan2025training, jain2025r2e}, unit tests remain more trustworthy provided that current language models and agents can generate adequate tests~\citep{chou2025autocodebenchlargelanguagemodels}. Second, the detailed checklist acts as a clear guide, steering the agent in implementing unit tests in the later stage.

Please refer to Appendix~\ref{subsection:prompttaskproposal} for the detailed prompt used for task proposal in our experiments.

\begin{figure*}[h!]
    \centering
    \includegraphics[width=1.0\textwidth]{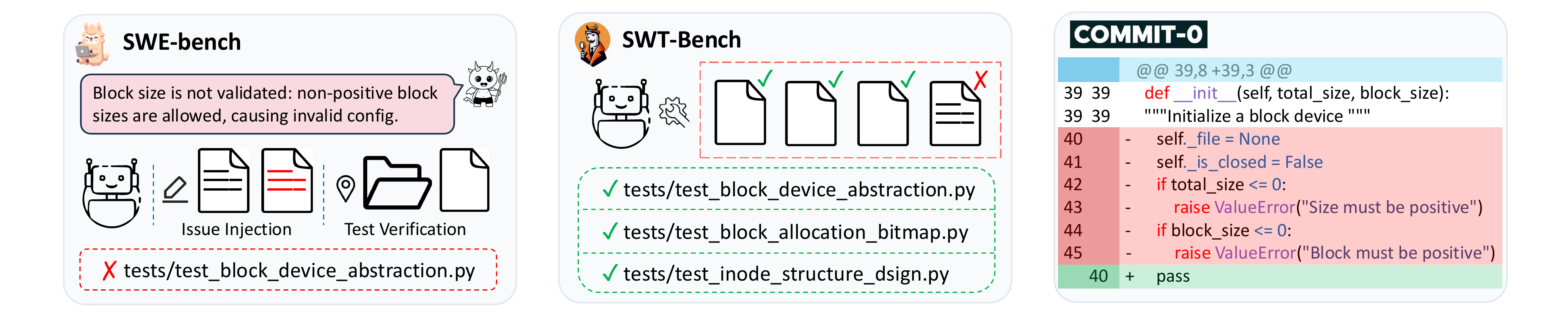}
    \vspace{-2em}
    \caption{Overview of capability-specific adaption as detailed in Section~\ref{subsection:taskspecificgeneration}. Our pipeline can be adopted for various coding tasks, such as issue resolution (SWE-bench), issue reproduction (SWT-Bench) and library generation from scratch (Commit-0) }
    \vspace{-0.5em}
    \label{fig:pipeine-2}
\end{figure*}

\subsection{Repository Setup}
\label{subsection:repositorysetup}

After the project proposal and step-by-step tasks are generated, the pipeline moves to the setup stage. An agent is in charge of establishing the foundational code structure, delineating necessary files, utilities, and function stubs without implementing any core functionality. Such practice ensures that all subsequent work is carried out within a predefined scaffolding, thereby confining the scope of implementation and preventing disorganized development.
Furthermore, we also request the agent to generate environmental dependencies and Docker related files, allowing for the creation of a readily executable sandboxed environment.

\subsection{Unit Test Generation}
\label{subsection:unittestgeneration}

Once the project scaffolding is in place, a dedicated agent\footnote{We use OpenHands CodeAct Agent in our experiments.} is responsible for generating the unit tests. The agent is guided by the checklist formulated during the task proposal stage, which ensures comprehensive test coverage and the implementation of all necessary tests. Operating as an agent, rather than a static LLM, provides the ability to access existing files and execute its own generated tests. This active verification capability ensures tests can correctly import dependencies, invoke functions, and produce expected errors. 

To standardize the evaluation pipeline across diverse projects, we require the agent to generate a uniform execution entry point for each task. This script orchestrates the environment setup and executes the unit tests, ensuring a consistent interface for the training loop. Furthermore, the generation process applies domain-specific verification criteria, such as enforcing numerical stability for mathematical tools, concurrency safety for database engines, or serialization compliance for network protocols.

In this stage, the mandate of our agent is to focus solely on the test requirements without anticipating how the functionality will be implemented. It is required to be maximally strict, compelling the subsequent implementation to conform precisely to these requirements to pass. This strictness is essential for preventing faulty implementations, even at the cost of increased implementation difficulty.

\subsection{Functionality Implementation}
\label{subsection:functionalityimplementation}

From project proposal to unit test generation, the pipeline prepares a coding project together with the unit tests, which collectively forms the environment for training coding agents. Following this, the agent, either trained or used for distillation, implements the core functionalities based on task instruction. In our implementation, the agent is given access to the unit test code. We find this necessary because as previously noted, the implementation of unit tests is highly rigorous and offers no tolerance for implementation deviations. Without this access, achieving a passing solution would be exceptionally difficult for the agent. It is worth noting that the adoption of such configuration is flexible, depending on both features of the generated unit tests and performance of the agent.

After the agent completes the implementation, the test suites in the repository are replaced with the original tests produced during the test generation stage, which is a critical step to mitigate the risk of reward hacking, where the agent might otherwise pass tests by directly editing the test code. The original unit tests are subsequently executed to verify the correctness of implementation. This overall process leverages a similar idea to the concept of self-questioning~\citep{chen2025selfquestioninglanguagemodels}, as unit test generation and functionality implementation form a mutual verification loop. An issue in either stage, be it a flawed test or an incorrect implementation, will lead to the failure of unit tests, thereby adding to trustworthiness of the generated unit tests.  and also robustness of the entire verification process. Notably, we alternate between unit test generation and functionality implementation, forming a loop. This approach ensures that subsequent unit tests are adapted to the the cumulative functionality, which simplifies the implementation the implementation process by removing the need of revisiting or modifying the previously validated code.

In our implementation, we mandate that the agent does not access the provided unit tests initially. Instead, it is encouraged to construct its own verification scripts based on the task description, a practice that simultaneously cultivates its test generation capabilities.

\begin{table*}[t]
\centering
\small
\begin{tabular}{l|c|cc|cc|c}
\toprule
\multirow{2}{*}[-1ex]{\textbf{Models}} & \multirow{2}{*}[-1ex]{\textbf{Dataset Size}} &
\multicolumn{2}{c|}{\textbf{SWE-bench Verified}} &
\multicolumn{2}{c|}{\textbf{SWT-Bench}} &
\multicolumn{1}{c}{\textbf{Commit-0}} \\
\cmidrule(lr){3-4} \cmidrule(lr){5-6} \cmidrule(lr){7-7}
& & Resolved Rate & Empty Patch Rate & Resolved Rate & Coverage Delta & Resolved Rate \\
\midrule\noalign{\vskip -3pt}
\multicolumn{7}{c}{\cellcolor[HTML]{EFEFEF} \scriptsize\textit{7B Models}} \\
\noalign{\vskip 3pt}
\textbf{Qwen2.5-Coder-7B} & - & \phantom{0}$1.8$ \phantom{\pos{0.00}} & $45.8$  \phantom{\pos{0.00}} & \phantom{0}$0.72$  \phantom{\pos{0.00}} & \phantom{0}$7.55$ \phantom{\pos{00.00}} & $1.82$  \phantom{\pos{0.00}} \\
\textbf{SWE-Gym-7B} & 491 & $10.6$ \pos{\phantom{0}8.8} & $33.8$ \negdrev{12.0} & \phantom{0}$1.45$ \pos{0.73} & \phantom{0}$5.19$ \negd{\phantom{0}2.36} & $2.51$ \pos{0.69} \\
\textbf{R2E-Gym-7B} & 3.3k & $19.0$ \pos{17.2} & -- & \phantom{0}$0.72$ \zerod & \phantom{0}$2.66$ \negd{\phantom{0}4.89} & $2.54$ \pos{0.72} \\
\textbf{SWE-smith-7B} & 2.0k & $15.2$ \pos{13.4} & -- & \phantom{0}$0.00$ \negd{0.72} & $12.30$ \pos{\phantom{0}4.75} & $2.62$ \pos{0.80} \\
\textbf{SWE-Play-mix-7B} & 704 & $17.0$ \pos{15.2} & \phantom{0}$6.4$ \negdrev{39.4} & \phantom{0}$3.26$ \pos{2.54} & $30.69$ \pos{23.14} & $2.95$ \pos{1.13} \\
\midrule\noalign{\vskip -3pt}
\multicolumn{7}{c}{\cellcolor[HTML]{EFEFEF} \scriptsize\textit{32B Models}} \\
\noalign{\vskip 3pt}
\textbf{Qwen2.5-Coder-32B} & - & \phantom{0}$7.0$ \phantom{\pos{0.00}} & \phantom{0}$9.5$ \phantom{\pos{0.00}} & \phantom{0}$9.42$ \phantom{\pos{0.00}} & $24.81$ \phantom{\pos{00.00}} & $2.31$ \phantom{\pos{0.00}} \\
\textbf{SWE-Gym-32B} & 491 & $20.6$ \pos{13.6} & $13.8$ \posrev{\phantom{0}4.3} & \phantom{0}$3.26$ \negd{6.16} & $10.34$ \negd{14.47} & $2.51$ \pos{0.20} \\
\textbf{R2E-Gym-32B} & 3.3k & $34.4$ \pos{27.4} & -- & \phantom{0}$3.26$ \negd{6.16} & $15.04$ \negd{\phantom{0}9.77} & $3.30$ \pos{0.99} \\
\textbf{SWE-smith-32B} & 5.0k & $40.2$ \pos{33.2} & -- & $13.77$ \pos{4.35} & $42.15$ \pos{17.34} & $2.62$ \pos{0.31} \\
\textbf{SWE-Play-mix-32B} & 704 & $31.2$ \pos{24.2} & \phantom{0}$6.8$ \negdrev{\phantom{0}2.7} & $18.12$ \pos{8.70} & $44.24$ \pos{19.43} & $3.64$ \pos{1.33} \\
\bottomrule
\end{tabular}
\vspace{-1em}
\caption{Model performance comparison of our SWE-Play-mix against the base model and baselines across three distinct coding benchmarks. For each result, we also report its performance delta relative to the corresponding Qwen2.5-Coder base model, with improvements highlighted in \textcolor{green!60!black}{green} and performance degradation in \textcolor{red!70!black}{red}.}
\vspace{-1em}
\label{tab:mainresults}
\end{table*}

\section{Capability-Specific Adaption}
\label{subsection:taskspecificgeneration}

The above pipeline supports the creation of general repository-level coding tasks. However, as discussed in Section~\ref{sec:codingskills}, software engineering requires solving a variety of tasks beyond this. To demonstrate the flexibility of \methodname, we showcase in this section how our pipeline can be adapted to generate tasks for three distinct use cases over which we measure performance in our experiments. Prompts can be found in Appendix~\ref{subsection:promptspecific}.

\textbf{Issue Resolution:}
SWE-bench~\citep{jimenez2024swebench} is the canonical benchmark for software engineering tasks, and it features resolving real GitHub issues. This task setting requires a repository that is generally functional but contains a minor bug affecting certain functionalities. With our pipeline, a functional repository already exists, with the only remaining task being the generation of a specific issue and its injection into the code. Similar to \citet{yang2025swesmith}, we first leverage an LLM to formulate an issue based on the unit test documentation and code of the task, and then call an agent to inject the issue into the repository. The existing test suites can directly verify the correctness of the injection, where a failing test serves as the success signal.

\textbf{Test Generation:}
SWT-Bench~\citep{mundler2024swtbench} is a test generation benchmark where agents are required to develop unit tests to reproduce issues. For this setting, beyond injecting the issue into the repository, we also require the agent to modify the test suites so that the injected bug is no longer directly detectable. The agents are then tasked with formulating a test script that exposes the faulty behavior and then integrate this script into the existing test suites to complete the task, a procedure identical to SWT-Bench.

\textbf{Generating Libraries from Scratch:}
Commit-0~\citep{zhao2024commit0librarygenerationscratch} evaluates agents on constructing the entire repository from scratch, with no step-to-step decomposition. For this setting, the complete and functional repositories generated by our pipeline serve as the target artifacts and reference implementation. To simulate the Commit-0 task format, we replace all function bodies with pass statements and then prompt agents to re-implement the project. Notably, we do not use the existing test suites to verify implementation correctness in our experiments. Instead, we let the agent attempt the full task and directly collect its trajectory, which aligns with a pure distillation setting. We adopt such strategy out of the practical necessity. Because our pipeline separates functionalities into step-by-step tasks and the unit tests are designed to be inherently stringent, it is challenging for the agent to pass all tests within a single call.

\section{Experiments}

\subsection{Experimental Settings}

With our proposed SWE-Playground pipeline, we collect from 28 generated projects a total of 280 general trajectories (as detailed in Section~\ref{subsection:projectproposal} to Section~\ref{subsection:functionalityimplementation}) and 424 task specific trajectories (as detailed in Section~\ref{subsection:taskspecificgeneration}), respectively 213 for issue resolution, 183 for issue reproduction and 28 for library generation. Following \citet{pan2025training} and \citet{jain2025r2e}, we use OpenHands~\citep{wang2025openhands} CodeAct Agent our agent Scaffold, and select Qwen-2.5-Coder series~\citep{hui2024qwen2} as our the base model. We adopt SWE-bench Verified~\citep{chowdhury2024swebenchverified}, SWT-Bench Lite and Commit-0 Lite as the evaluation splits for the respective benchmarks. We provide details in Appendix~\ref{appendix:expdetail}.

\label{subsection:experimentsettings}

\begin{table*}[t]
\centering
\begin{tabular}{@{\hspace{0.1cm}}l|cc|cc|c}
\toprule
\multirow{2}{*}[-1ex]{\textbf{Models}} &
\multicolumn{2}{c|}{\textbf{SWE-bench Verified}} &
\multicolumn{2}{c|}{\textbf{SWT-Bench}} &
\multicolumn{1}{c}{\textbf{Commit-0}} \\
\cmidrule(lr){2-3} \cmidrule(lr){4-5} \cmidrule(lr){6-6}
& Resolved Rate & Empty Patch Rate & Resolved Rate & Coverage Delta & Resolved Rate \\
\midrule
\textbf{Qwen2.5-Coder} & \phantom{0}$1.8$ & $45.8$ & $0.72$ & \phantom{0}$7.55$ & $1.82$ \\
\textbf{SWE-Play-mix} & $17.0$ & \phantom{0}$6.4$ & $3.26$ & $30.69$ & $2.95$ \\
\textbf{SWE-Play-general} & \phantom{0}$8.6$ & $22.6$ & $2.54$ & \phantom{0}$7.55$ & $2.62$ \\
\textbf{SWE-Play-swe} & $11.0$ & $10.6$ & $0.72$ & \phantom{0}$4.80$ & $2.38$ \\
\bottomrule
\end{tabular}
\vspace{-0.5em}
\caption{Performance comparison of different SWE-Play 7B models on all the evaluated benchmarks.}
\vspace{-1em}
\label{tab:comparison}
\end{table*}

\subsection{Results and Analysis}

\textbf{\methodname trains agents with strong performance across various benchmarks.} As the results show, SWE-Play-mix-7B, trained with trajectories generated by our proposed \methodname, consistently surpasses its base model across three distinct benchmarks. Notably, it secures top performance on four of five evaluation metrics against strong baselines and is only marginally outperformed by R2E-Gym-7B on SWE-bench Verified. Similarly, SWE-Play-mix-32B exhibits a comparable improvement trend, achieving top performance on SWT-Bench, Commit-0, and the SWE-bench Verified empty patch rate. This result is particularly compelling given that our model is trained on a substantially smaller dataset, which underscores the exceptional data efficiency of our framework.

\textbf{Models trained with prior methods do not generalize well.} The results reveal the limited generalization of existing environments. While agents trained with thse environments perform competitively on SWE-bench, their capabilities falter on SWT-Bench. On this benchmark, 7B models show mixed results: only SWE-Gym-7B improves the resolved rate and SWE-smith-7B improves the coverage delta. In contrast, for 32B models, both SWE-Gym-32B and R2E-Gym-32B demonstrate performance degradation. Although SWE-smith achieves decent improvement, it is surpassed by our model, particularly on the resolved rate, which evaluates whether the model truly solved the instance. Similarly, on Commit-0, these methods lag considerably behind our models despite showing some improvement over the baseline. These findings indicate these training pipelines encourage agents to overfit to a narrow task distribution, hindering their development of versatile coding skills. Conversely, the varied nature and task adaptation strategy of the trajectories generated by \methodname can foster stronger cross-task performance, enabling agents to be more robust and generalizable. As we have argued, all these coding benchmarks hold equal importance for real-world application, and a truly capable coding agent should be competitive across the entire suite of tasks.

\textbf{7B models and 32B models perform inconsistently.}
According to our result, 7B and 32B models exhibit distinct performance patterns. Firstly, on SWE-bench Verified, models trained with SWE-Gym, R2E-Gym, and our SWE-Play-mix show similar improvement trends on the resolved rate. The improvement of 32B models is consistently around 1.6 times that of 7B models, with the only exception being SWE-smith, which uses significantly more trajectories for training its 32B model. On the empty patch rate for 32B models, the base model is already highly competitive, causing SWE-Gym-32B to ultimately underperform it. However, even under this circumstance, our SWE-Play-32B still achieves a lower empty patch rate, demonstrating the robustness of our collected data. Secondly, on SWT-Bench, though SWE-Gym and R2E-Gym both exhibit comparable results to the baseline for 7B models, their performance consistently degrades for 32B models. This is likely because the Qwen2.5-Coder-7B is already mostly incapable of these tasks, meaning its performance is already at a floor and cannot be degraded further. However, for the more capable Qwen2.5-Coder-32B, this performance gain is lost, demonstrating that these methods are inherently detrimental to model capability of reproducing issues as evaluated by SWT-Bench. However, SWE-smith models show a completely different pattern on this benchmark. SWE-smith-7B, which achieves a higher coverage delta while failing to solve any instances, is far surpassed by the performance of SWE-smith-32B which demonstrates a strong improvement.

\section{Ablations and Discussions}

\subsection{Impact of Trajectory Composition}
\label{subsection:trajectory}

To investigate how different trajectory types influence model performance, we conduct an ablation study by training another two 7B model: SWE-Play-general with 280 general-purpose trajectories, and SWE-Play-swe with 213 issue resolution trajectories, as described in Section~\ref{subsection:experimentsettings}. The results are presented in Table~\ref{tab:comparison}.

Results indicate that training solely on general trajectories yields performance gains on three of the four metrics. The only exception is a nearly same coverage delta on SWT-Bench. Such finding demonstrates that generalist training, which focuses on implementing functionalities within large code repositories, effectively develops core coding abilities for the agents. We attribute this success to the intrinsic design of the \methodname task proposal and data collection pipeline. Firstly, in our tasks, models are required to build projects from a blank template, a process analogous to generating a library from the initial commit of Commit-0. Secondly, the natural process of implementing new functionality often involves identifying and resolving bugs, which directly enhances model performance on issue resolution tasks in SWE-bench. Finally, during data collection, we explicitly prompt the model to write its own test scripts before implementing the core functionalities or accessing provided test files. This practice empowers the model with the skills of unit test generation and issue reproduction. Consequently, though not identical in prompt format and task type to SWE-bench, SWT-Bench, or Commit-0, the general trajectories still result in substantial performance improvement.

Furthermore, as expected, training with issue resolution trajectories boosts model performance on SWE-bench Verified. While SWE-Play-swe slightly outperforms SWE-Play-general on this specific benchmark, it lags considerably behind SWE-Play-mix, which is trained on a combination of all collected trajectories, containing all data used for SWE-Play-swe, plus the general trajectories and additional trajectories tailored for issue reproduction and library generation. The superior performance of SWE-Play-mix highlights the importance of data diversity.

Given that general and issue resolution trajectories are known to boost performance, we further seek to validate the contribution of our other two types of trajectories: issue reproduction and library generation. We hypothesize that these trajectories provide complementary skills, leading to a synergistic improvement. To test this, we train an ablation model using only a mixture of general and issue resolution trajectories, and results are provided in Table~\ref{tab:further}. This ablated model achieves the same resolved rate on SWE-bench Verified as the model trained only with issue resolution trajectories, while both significantly outperformed by our full SWE-Play-mix model. This gap demonstrates that the inclusion of issue reproduction and library generation trajectories is the critical factor that contributes to the further improvement, proving their direct and positive influence on the issue resolution capability and confirming the effectiveness of cross-scenario transferability.

\begin{table}[h]
    \centering
    \small
    \begin{tabular}{@{\hspace{0.1cm}}l|c|c@{\hspace{0.2cm}}c}
        \toprule
        \multirow{2}{*}[-1ex]{\textbf{Traj. Comp.}} & \multirow{2}{*}[-1ex]{\textbf{Num.}} & \multicolumn{2}{c}{\textbf{SWE-bench Verified}} \\
        \cmidrule(lr){3-4}
        & & \textbf{Resolved Rate} & \textbf{Empty Patch Rate} \\
        \midrule
        \textbf{General} & 280 & \phantom{0}8.6 & 22.6 \\
        \textbf{SWE-bench} & 213 & 11.0 & 10.6 \\
        \textbf{Gen \& SWE} & 493 & 11.0 & 12.6 \\
        \textbf{Full Mixed} & 704 & 17.0 & \phantom{0}6.4 \\
        \bottomrule
    \end{tabular}
    \vspace{-1em}
    \caption{Model Performance on SWE-bench Verified with four distinct trajectory compositions: pure general trajectories, pure issue resolution trajectories, general trajectories plus issue resolution trajectories, and full mixture of all collected trajectories.}
    \label{tab:further}
\end{table}

\subsection{Investigation into Data Efficiency}

Results indicate that models trained on \methodname achieve similar comparable performance with fewer trajectories. Within all the evaluated benchmarks for 7B models, SWE-Play-mix is only outperformed by R2E-Gym on SWE-bench Verified, which relies on a training dataset nearly five times larger (3,321 trajectories versus our 704). Moreover, as specified in Section~\ref{subsection:trajectory}, our SWE-Play-swe-7B surpasses SWE-Gym-7B on SWE-bench Verified while using less than half the volume of identically formatted training trajectories.

Crucially, this efficiency advantage persists when accounting for the total tokens of training data. As detailed in Table~\ref{tab:trajectory}, while our trajectories contain more tokens on average, the aggregate size of our dataset ($\sim$27.5M tokens) remains significantly smaller than that of R2E-Gym ($\sim$45.8M tokens). This confirms that \methodname achieves competitive performance with reduced computational overhead, regardless of whether the cost is measured by instance count or total token consumption. 

These outcomes raise a critical question: why can our framework demonstrate better data efficiency compared with various previous methods? To answer this, we perform a detailed comparative analysis of our trajectories against those from baseline studies, with the results presented in Table~\ref{tab:trajectory}. Our analysis reveals that trajectories generated with \methodname are quantifiably more complex and comprehensive. As shown in the table, they contain two to three times more tokens and messages on average, including a higher number of assistant tokens, tool calls and lines of code edited, compared to other datasets. We characterize this quality as a higher learning density. This concept, in line with the Agency Efficiency Principle introduced by \citet{xiao2025limiagency}, suggests that trajectories dense with diverse and complex actions provide a more potent learning impetus per sample. The richness of our trajectories demonstrates this principle in practice, leading to greater data efficiency.

Furthermore, our trajectories exhibit a significantly higher proportion of bash execution actions. This indicates that the agent learns a more robust development methodology, dedicating more steps to inspecting the project structure and verifying its implementation rather than focusing narrowly on code generation. We posit that such emphasis on a realistic iterative workflow provides a more pragmatic learning foundation and thus yields a more capable agent.

\begin{table}[h]
    \centering
    \small
    \begin{tabular}{@{\hspace{0.1cm}}c|c@{\hspace{0.3cm}}c@{\hspace{0.3cm}}c@{\hspace{0.1cm}}}
        \toprule
        \textbf{Metric} & \textbf{SWE-Gym} & \textbf{R2E-Gym} & \textbf{SWE-Play-mix} \\
        \midrule
        Total Count & 491 & 3.3k  & 704 \\
        Message & 40.2 & 33.2 & 72.9 \\
        Tokens & 18,377 & 13,795 & 39,121 \\
        Assist Tokens & 2,595  & 2,604  & 7,308 \\
        Tool Call & 17.9  & 16.1  & 36.0 \\
        Lines Edited & 17.5  & 38.6  & 103.5 \\
        Bash Proportion & 27.8\%  & 26.7\%  & 41.7\% \\
        \bottomrule
    \end{tabular}
    \vspace{-0.3em}
    \caption{Trajectory statistics for different datasets. We report the total number of trajectories \textbf{(Total Count)} and several metrics averaged per trajectory: number of messages \textbf{(Message)}, total tokens \textbf{(Token)}, assistant tokens \textbf{(Assist Token)}, tool call trials \textbf{(Tool Call)}, and edited code lines \textbf{(Lines Edited)}. \textbf{Bash Proportion} is the percentage of bash executions among all tool calls.}
    \vspace{-0.5em}
    \label{tab:trajectory}
\end{table}

\section{Conclusion and Future Work}

In this paper, we present \methodname, a novel environment for training versatile coding agents. Our results demonstrate that \methodname supports training agents with strong performance across various coding benchmarks with a remarkably smaller dataset. We holistically argue that performance improvement on SWE-bench alone is insufficient, and proficiency across multiple benchmarks is critical for training coding agents.

While these results are promising, we identify two key directions for future work. Firstly, incorporating a broader set of benchmarks, such as SWE-bench Multimodal and SWE-perf, would further validate the extensibility of our framework, as their features align well with the generative capabilities of \methodname pipeline. Secondly, conducting RL experiments is an interesting direction to explore the full potential of our environment generation pipeline. As we have argued, our pipeline creates a scenario where unit test generation and code implementation mutually verify each other. Training an agent to master both skills through RL could directly probe its capacity to develop self-verifying and, ultimately, self-improving coding agents like \citet{liu2025agent0vl}.

\section*{Impact Statement}

In this paper, we present \methodname, an automated pipeline to synthesize environments and trajectories for the training of versatile coding agents. We anticipate several potential impacts of this work. First, \methodname offers a pathway to advance the field of automated software engineering. As discussed, real-world software engineering requires a diverse set of skills beyond simple code completion, and \methodname offers a robust framework for incorporating and training these versatile capabilities. From a broader perspective, with more research attention focused on the purely synthetic training data and self-evolving agents, \methodname shows promising results for further research regarding these topics.

At the same time, although our experiments demonstrate the effectiveness of the \methodname pipeline, rigorous quality control is essential if these systems are to be deployed in high-stakes scenarios. Furthermore, generating unit tests with LLMs remains an emerging research topic. Currently, there is a scarcity of benchmarks designed to evaluate this specific capability, and few works have discussed it systematically. Consequently, further research is required before models can be fully and reliably incorporated for automated verification.

\bibliography{example_paper}
\bibliographystyle{icml2025}

\newpage
\appendix
\onecolumn

\section{Related Work}
\label{sec:existingbenchmarks}

\textbf{Coding Agents.} Capable Large Language Models and robust agent scaffolds are two critical components of modern coding agents. Recent state-of-the-art language models include proprietary ones such as Claude Sonnet 4.5~\citep{anthropic2025claude45} and GPT-5~\citep{openai2025gpt5}, as well as open-source models such as Qwen3-Coder~\citep{qwen3technicalreport}, Kimi K2~\citep{kimiteam2025kimik2openagentic} and DeepSeek-V3.1~\citep{deepseekai2024deepseekv3technicalreport}. For coding agents scaffolds, SWE-agent~\citep{yang2024sweagent} introduces the concept of agent-computer interface which offers a small set of simple actions for interacting with files. Building upon SWE-agent, mini-SWE-agent~\citep{yang2024sweagent} achieves similar performance with a simpler control flow. OpenHands~\citep{wang2025openhands} expands on this by offering a wider action space including web browsing and API calls, enabling models to tackle more complex and diverse tasks. SWE-search ~\citep{antoniades2024swesearchenhancingsoftwareagents} employs a multi-agent system and Monte Carlo Tree Search (MCTS) strategy for planning to enhance flexibility during state transition. The combination of these powerful models and agent scaffolds enables coding agents to solve real-world coding challenges and even surpass the performance of human engineers on certain tasks~\citep{li2025aiteammates} .

\textbf{Coding Tasks and Benchmarks.} HumanEval~\citep{chen2021codex} is the first coding benchmark to use an executable environment for verification instead of relying on vague text similarity, laying the foundation for subsequent work. LiveCodeBench~\citep{jain2024livecodebench} advances this by increasing the difficulty and evaluation coverage, gathering problems from contests across various competition platforms. BigCodeBench~\citep{zhuo2024bigcodebench} challenges models to follow complex instructions to invoke function calls as tools for solving practical tasks. In addition to these function-level and file-level edition tasks, a large number of recent benchmarks have been designed to test the performance of models and agents on repository-level coding tasks. Commit-0~\citep{zhao2024commit0librarygenerationscratch} targets rebuilding existing libraries from scratch. SWE-bench~\citep{jimenez2024swebench} is also crafted from real GitHub repositories and challenges models to solve issues noted in pull requests. SWE-Bench Pro~\citep{deng2025swebenchproaiagents} builds upon SWE-bench with more challenging tasks to capture realistic, complex, enterprise-level problems beyond the scope of SWE-Bench. SWE-bench Multimodal~\citep{yang2024swebenchmultimodal} and SWE-bench Multilingual~\citep{yang2025swesmith} both extend SWE-bench to evaluate more diverse abilities. SWE-bench Multimodal incorporates visual elements such as diagrams and screenshots to test whether modals can perform well on visual information. SWE-bench Multilingual covers nine different programming languages for the evaluation of various application domains. Multi-SWE-bench~\citep{zan2025multiswebench} is also a issue-resolving benchmark aiming to evaluate various programming languages. SWT-Bench~\citep{mundler2024swtbench} evaluates model proficiency on reproducing GitHub issues in Python. TestGenEval~\citep{jain2025testgenevalrealworldunit} measures test generation performance in well-maintained Python repositories. ArtifactsBench~\citep{zhang2025artifactsbenchbridgingvisualinteractivegap} assesses the ability to transform multimodal instructions into high-quality and interactive visual artifacts. SWE-Perf~\citep{he2025sweperf} and SWE-fficiency~\citep{ma2025swefficiencylanguagemodelsoptimize} benchmarks models on optimizing code performance in real-world repositories. KernelBench~\citep{ouyang2025kernelbenchllmswriteefficient} evaluates model capability on writing efficient GPU kernels. We argue that \textbf{all the evaluated abilities are equally important for coding agents, and the exceptional performance on a single benchmark is not indicative of a truly capable agent}.

We provide a comprehensive review of these representative coding tasks and benchmarks in Table~\ref{tab:relatedwork}.

\begin{table}[h]
\centering
\scriptsize
\begin{tabular}{l|c|c|c|c|c}
\toprule
Benchmark & Released & Count & Interaction & Focus & Construction \\
\midrule
HumanEval~\citep{chen2021codex} & 2021.7\phantom{0} & 164 & Static & Fuction-Level Functional Correctness & Human Crafted \\
LiveCodeBench~\citep{jain2024livecodebench} & 2024.3\phantom{0} & 511 & Static & Holistic Coding Abilities on Contest Problems & Auto Collected \\
BigCodeBench~\citep{zhuo2024bigcodebench} & 2024.6\phantom{0} & 1,140 & Static & Complex Instruction Following with API Calling & Human Crafted \\
Commit-0~\citep{zhao2024commit0librarygenerationscratch} & 2024.9\phantom{0} & 54 & Agentic & Repository-Level Library Generation From Scratch & Auto Collected \\
SWE-bench~\citep{jimenez2024swebench} & 2023.10 & 2,294 & Agentic & GitHub Issue Resolution & Auto Collected \\
SWE-bench Pro~\citep{deng2025swebenchproaiagents} & 2025.9\phantom{0} & 1,865 & Agentic & Long-Horizon Enterprise-Level Multi-File Patches & Mixed Curation \\
SWE-bench Multimodal~\citep{yang2024swebenchmultimodal} & 2024.10 & 517 & Agentic & Visual Context Understanding \& Generation & Auto Collected \\
SWE-bench Multilingual~\citep{yang2025swesmith} & 2025.5\phantom{0} & 300 & Agentic & Multiple Programming Languages & Auto Collected \\
Multi-SWE-bench~\citep{zan2025multiswebench} & 2025.4\phantom{0} & 4,723 & Agentic & Multilingual Issue Resolution & Auto Collected \\
SWT-Bench~\citep{mundler2024swtbench} & 2024.6\phantom{0} & 1,900 & Agentic & Test Generation for Bug Reproduction & Auto Collected \\
TestGenEval~\citep{jain2025testgenevalrealworldunit} & 2024.10 & 1,210 & Static & Paired Test Generation \& Completion & Human Crafted  \\
ArtifactsBench~\citep{zhang2025artifactsbenchbridgingvisualinteractivegap} & 2025.7\phantom{0} & 1,825 & Static & Interactive Visual Artifacts & Human Crafted \\
SWE-Perf~\citep{he2025sweperf} & 2025.7\phantom{0} & 140 & Agentic & Code Performance Optimization & Auto Collected \\
SWE-fficiency~\citep{ma2025swefficiencylanguagemodelsoptimize} & 2025.11 & 498 & Agentic & Repository-Level Performance Optimization & Human Crafted  \\
KernelBench~\citep{ouyang2025kernelbenchllmswriteefficient} & 2025.2\phantom{0} & 250 & Static & Efficient PyTorch Kernel Generation & Human Crafted \\
\bottomrule
\end{tabular}
\caption{Summary of representative coding benchmarks, including release time, count of instances, interaction type, major evaluated capability, and the way of construction.}
\label{tab:relatedwork}
\end{table}

\section{Experimental Details}

\label{appendix:expdetail}

\subsection{Training Details}

\textbf{Agent Scaffold.} Following \citet{pan2025training} and \citet{jain2025r2e}, we use OpenHands~\citep{wang2025openhands} CodeAct Agent as our agent scaffold. OpenHands CodeAct Agent leverages ReAct framework~\citep{yao2023react}, empowering agents with the ability to interact with the command line, edit files and browse the web based on a reasoning and acting loop. OpenHands also provides securely isolated Docker sandboxes for all the actions to be executed, thereby ensuring a safe and consistent execution environment that prevents unintended changes to the host system. In our experiments, we use version 0.48.0\footnote{https://github.com/All-Hands-AI/OpenHands/releases/tag/0.48.0} of OpenHands and disable the browsing tool and function call features.

\textbf{Dataset Collection and Statistics.} We employ an ensemble of Gemini 2.5 Pro~\citep{comanici2025gemini25pushingfrontier}, GPT-4.1~\citep{openai2025gpt41} and Claude Sonnet 4~\citep{anthropic2025claude4} for proposing projects. For proposing step-by-step tasks, we use GPT-4.1. All agents involved in our pipeline are implemented using OpenHands in conjunction with Claude Sonnet 4. We collect from 28 generated projects a total of 280 general trajectories (as detailed in Section~\ref{subsection:projectproposal} to Section~\ref{subsection:functionalityimplementation}) and 424 capability-specific trajectories (as detailed in Section~\ref{subsection:taskspecificgeneration}), respectively 213 for issue resolution, 183 for issue reproduction and 28 for library generation, with Claude Sonnet 4. In experiments, we exclusively use the trajectories of finishing the task instead of those for generating the task. While the trajectories for unit test generation and issue injection could be valuable for training agents on specialized tasks such as generating unit tests from documentation, we do not include such benchmarks in our experiments, with the exception of SWT-Bench~\citep{mundler2024swtbench}, which requires agents to reproduce issues, but not to purely write unit tests. Therefore, we only adopt the trajectories that were are relevant to the capabilities evaluated in the benchmarks, namely, the trajectories of building up the project, fixing bugs and reproducing issues.

\textbf{Training details.} Following previous works~\citep{pan2025training, jain2025r2e, yang2025swesmith}, we select Qwen-2.5-Coder series~\citep{hui2024qwen2}, Qwen-2.5-Coder-7B-Instruct and Qwen-2.5-Coder-32B-Instruct, as our base models. We finetune the base model with all the trajectories mixed together and randomly shuffled, using a learning rate of \texttt{1e-5} and training for 3 epochs. Given that the length of our functionality implementation trajectories frequently exceeds the default maximum context length of \texttt{32,768} tokens, we follow the official instructions\footnote{https://huggingface.co/Qwen/Qwen2.5-Coder-7B-Instruct} to extend context length using YaRN~\citep{peng2024yarn} during both training and inference stage. Following \citet{jain2025r2e}, we use LLaMA-Factory~~\citep{zheng2024llamafactory} for efficient model finetuning, and adopt \citet{zou360lf} to further integrate sequence parallelism feature.

\textbf{Benchmarks and Baselines.} We select three distinct benchmarks to demonstrate the effectiveness of \methodname: SWE-bench, SWT-Bench and Commit-0. SWE-bench~\citep{jimenez2024swebench} focuses on solving real GitHub issues. SWT-Bench~\citep{mundler2024swtbench} evaluates agents on their ability to reproduce issues. Commit-0~\citep{zhao2024commit0librarygenerationscratch} tasks agents with building Python libraries from scratch. Collectively, these three benchmarks cover a wide range of critical coding capabilities, allowing for a comprehensive understanding on whether agents can handle various tasks. We adopt SWE-bench Verified~\citep{chowdhury2024swebenchverified}, SWT-Bench Lite and Commit-0 Lite as the evaluation splits for the respective benchmarks. We report resolved rate and empty patch rate for SWE-bench Verified, resolved rate and coverage delta for SWT-Bench Lite, and resolved rate for Commit-0 Lite. We select three strong baselines that focus on constructing coding environments and training SWE agents: SWE-Gym~\citep{pan2025training}, R2E-Gym~\citep{jain2025r2e} and SWE-smith~\citep{yang2025swesmith}. Similar to our approach, SWE-Gym and R2E-Gym also build upon OpenHands, while SWE-smith is based on SWE-agent framework~\citep{yang2024sweagent}. We use vLLM~\citep{kwon2023efficient} to serve all the models for inference. We use OpenHands remote runtime~\citep{neubig2024-evaluation-of-llms} to host docker containers during evaluation.


\subsection{Evaluation Prompts}

We provide our prompts used in our experiments for evaluating SWE-bench, SWT-Bench and Commit-0 with OpenHands framework as follows.

\begin{tcolorbox}[breakable, title=Prompt for SWE-bench]
\small
\textless uploaded\_files\textgreater \\
/workspace/\{\{ workspace\_dir\_name \}\}\\
\textless /uploaded\_files\textgreater \\

I've uploaded a python code repository in the directory \{\{ workspace\_dir\_name \}\}. Consider the following issue description:\\

\textless issue\_description\textgreater \\
\{\{ instance.problem\_statement \}\}\\
\textless /issue\_description\textgreater \\

Can you help me implement the necessary changes to the repository so that the requirements specified in the \textless issue\_description\textgreater  are met?\\
I've already taken care of all changes to any of the test files described in the \textless issue\_description\textgreater . This means you DON'T have to modify the testing logic or any of the tests in any way!\\
Also the development Python environment is already set up for you (i.e., all dependencies already installed), so you don't need to install other packages.\\
Your task is to make the minimal changes to non-test files in the /workspace/\{\{ workspace\_dir\_name \}\} directory to ensure the \textless issue\_description\textgreater  is satisfied.\\

Follow these phases to resolve the issue:\\

Phase 1. READING: read the problem and reword it in clearer terms\\
   1.1 If there are code or config snippets. Express in words any best practices or conventions in them.\\
   1.2 Hightlight message errors, method names, variables, file names, stack traces, and technical details.\\
   1.3 Explain the problem in clear terms.\\
   1.4 Enumerate the steps to reproduce the problem.\\
   1.5 Hightlight any best practices to take into account when testing and fixing the issue\\

Phase 2. RUNNING: install and run the tests on the repository\\
   2.1 Follow the readme\\
   2.2 Install the environment and anything needed\\
   2.2 Iterate and figure out how to run the tests\\

Phase 3. EXPLORATION: find the files that are related to the problem and possible solutions\\
   3.1 Use \textasciigrave grep\textasciigrave  to search for relevant methods, classes, keywords and error messages.\\
   3.2 Identify all files related to the problem statement.\\
   3.3 Propose the methods and files to fix the issue and explain why.\\
   3.4 From the possible file locations, select the most likely location to fix the issue.\\

Phase 4. FIX ANALYSIS: state clearly the problem and how to fix it\\
   4.1 State clearly what the problem is.\\
   4.2 State clearly where the problem is located.\\
   4.3 State clearly how the test reproduces the issue.\\
   4.4 State clearly the best practices to take into account in the fix.\\
   4.5 State clearly how to fix the problem.\\

Phase 5. FIX IMPLEMENTATION: Edit the source code to implement your chosen solution.\\
   5.1 Make minimal, focused changes to fix the issue.\\

Phase 6. VERIFICATION: Test your implementation thoroughly.\\
   6.1 Run existing tests related to the modified code to ensure you haven't broken anything.\\

Phase 7. FINAL REVIEW: Carefully re-read the problem description and compare your changes with the base commit \{\{ instance.base\_commit \}\}.\\
   7.1 Ensure you've fully addressed all requirements.\\
   7.2 Run any tests in the repository related to:\\
     7.2.1 The issue you are fixing\\
     7.2.2 The files you modified\\
     7.2.3 The functions you changed\\
   7.3 If any tests fail, revise your implementation until all tests pass\\

Be thorough in your exploration, testing, and reasoning. It's fine if your thinking process is lengthy - quality and completeness are more important than brevity.\\

Do not run the provided test files to verify your fix, as the issue can not be detected by the test files.\\
\end{tcolorbox}

\begin{tcolorbox}[breakable, title=Prompt for SWT-Bench]
\small
\textless uploaded\_files\textgreater \\
/workspace/{{ workspace\_dir\_name }}\\
\textless /uploaded\_files\textgreater \\
I've uploaded a python code repository in the directory \{\{ workspace\_dir\_name \}\}. Consider the following issue description:\\

\textless issue\_description\textgreater \\
\{\{ instance.problem\_statement \}\}\\
\textless /issue\_description\textgreater \\

Can you help me implement the necessary changes to the repository to test whether the issue in \textless issue\_description\textgreater  was resolved?\\
I will take care of all changes to any of the non-test files. This means you DON'T have to modify the actual logic and ONLY have to update test logic and tests!\\
Your task is to make the minimal changes to tests files in the /workspace directory to reproduce the issue in the \textless issue\_description\textgreater , i.e., such that the generated tests fail in the current state (where the issue is unresolved) and pass when the issue will be resolved.\\
Follow these steps to reproduce the issue:\\
1. As a first step, it might be a good idea to explore the repo to familiarize yourself with its structure.\\
2. Create a script \textasciigrave reproduction.py\textasciigrave  to reproduce the error and execute it with \textasciigrave python reproduction.py\textasciigrave  using the BashTool, to confirm the error\\
3. Edit the sourcecode of the repo to integrate your reproduction script into the test framework\\
4. Run the test framework and make sure your tests fail! Only submit FAILING tests! Never submit passing tests.\\
\{\{ test\_instructions \}\}Your thinking should be thorough and so it's fine if it's very long.\\
\end{tcolorbox}

\begin{tcolorbox}[breakable, title=Prompt for Commit-0]
\small
\textless uploaded\_files\textgreater \\
/workspace/\{\{ workspace\_dir\_name \}\}\\
\textless /uploaded\_files\textgreater \\
I've uploaded a python code repository in the directory \{\{ workspace\_dir\_name \}\}. Here is your task:\\

Here is your task:\\

  You need to complete the implementations for all functions (i.e., those with pass statements) and pass the unit tests.\\

  Do not change the names of existing functions or classes, as they may be referenced from other code like unit tests, etc.\\

  When you generate code, you must maintain the original formatting of the function stubs (such as whitespaces), otherwise we will not able to search/replace blocks for code modifications, and therefore you will receive a score of 0 for your generated code.\\

Here is the command to run the unit tests:\\
\textless test\_command\textgreater \\
pytest .\\
\textless /test\_command\textgreater \\

Make a local git commit for each agent step for all code changes. If there is not change in current step, do not make a commit.\\
\end{tcolorbox}

\section{Extended Results}

We provide our full experimental results in Table~\ref{tab:fullresults}. In addition to SWE-bench, SWT-Bench and Commit-0, we also evaluate models on BigCodeBench~\citep{zhuo2024bigcodebench}, a \textbf{non-agentic} benchmark which assesses models on programming challenges. As the results show, all evaluated models demonstrate performance degradation against the base model on this benchmark, with SWE-smith-7B performing the worst with only a $6.08\%$ success rate. This result validates that training with agentic trajectories inevitably comes at the cost of the capability on non-agentic static tasks. Firstly, models trained with agentic tasks typically adapt to the complex interactions and execution-based verification nature of agentic environments, and thus falter when facing static coding tasks. Secondly, agentic tasks such as SWE-bench are inherently different in format and capabilities evaluated from non-agentic tasks such as BigCodeBench, with the former focusing on context and project understanding, while the latter stressing algorithm design and instruction following. Furthermore, we hypothesize that this effect might be largely influenced by the design of the agent scaffold used during training.

\begin{table*}[h]
\centering
\scriptsize
\begin{tabular}{l|c|cc|cc|c|c}
\toprule
\multirow{2}{*}[-1ex]{\textbf{Models}} & \multirow{2}{*}[-1ex]{\textbf{Dataset Size}} &
\multicolumn{2}{c|}{\textbf{SWE-bench Verified}} &
\multicolumn{2}{c|}{\textbf{SWT-Bench}} &
\multicolumn{1}{c|}{\textbf{Commit-0}} &
\multicolumn{1}{c}{\textbf{BigCodeBench}} \\
\cmidrule(lr){3-4} \cmidrule(lr){5-6} \cmidrule(lr){7-7} \cmidrule(lr){8-8}
& & Resolved Rate & Empty Patch Rate & Resolved Rate & Coverage Delta & Resolved Rate & Success Rate \\
\midrule\noalign{\vskip -3pt}
\multicolumn{8}{c}{\cellcolor[HTML]{EFEFEF} \scriptsize\textit{7B Models}} \\
\noalign{\vskip 3pt}
\textbf{Qwen2.5-Coder-7B} & - & \phantom{0}$1.8$ & $45.8$ & \phantom{0}$0.72$ & \phantom{0}$7.55$ & $1.82$ & $17.58$ \\
\textbf{SWE-Gym-7B} & 491 & $10.6$ & $33.8$ & \phantom{0}$1.45$ & \phantom{0}$5.19$ & $2.51$ & $12.16$ \\
\textbf{R2E-Gym-7B} & 3.3k & $19.0$ & -- & \phantom{0}$0.72$ & \phantom{0}$2.66$ & $2.54$ & $16.90$ \\
\textbf{SWE-smith-7B} & 2.0k & $15.2$ & -- & \phantom{0}$0.00$ & $12.30$ & $2.62$ & \phantom{0}$6.08$ \\
\textbf{SWE-Play-mix-7B} & 704 & $17.0$ & \phantom{0}$6.4$ & \phantom{0}$3.26$ & $30.69$ & $2.95$ & $14.19$ \\
\textbf{SWE-Play-general-7B} & 280 & \phantom{0}$8.6$ & $22.6$ & \phantom{0}$2.54$ & \phantom{0}$7.55$ & $2.62$ & $14.86$ \\
\textbf{SWE-Play-swe-7B} & 213 & $11.0$ & $10.6$ & \phantom{0}$0.72$ & \phantom{0}$4.80$ & $2.38$ & -- \\
\textbf{SWE-Play-genswe-7B} & 493 & $11.0$ & $12.6$ & -- & -- & -- & -- \\
\midrule\noalign{\vskip -3pt}
\multicolumn{8}{c}{\cellcolor[HTML]{EFEFEF} \scriptsize\textit{32B Models}} \\
\noalign{\vskip 3pt}
\textbf{Qwen2.5-Coder-32B} & - & \phantom{0}$7.0$ & \phantom{0}$9.5$ & \phantom{0}$9.42$ & $24.81$ & $2.31$ & -- \\
\textbf{SWE-Gym-32B} & 491 & $20.6$ & $13.8$ & \phantom{0}$3.26$ & $10.34$ & $2.51$ & -- \\
\textbf{R2E-Gym-32B} & 3.3k & $34.4$ & -- & \phantom{0}$3.26$ & $15.04$ & $3.30$ & -- \\
\textbf{SWE-smith-32B} & 5.0k & $40.2$ & -- & $13.77$ & $42.15$ & $2.62$ & -- \\
\textbf{SWE-Play-mix-32B} & 704 & $31.2$ & \phantom{0}$6.8$ & $18.12$ & $44.24$ & $3.64$ & -- \\
\textbf{SWE-Play-swe-32B} & 213 & $29.4$ & \phantom{0}$6.8$ & -- & -- & -- & -- \\
\bottomrule
\end{tabular}
\caption{Full results of our experiments for all models, including the results on BigCodeBench.}
\label{tab:fullresults}
\end{table*}

\section{Prompt Details}

We provide in this section the detailed prompts used for the SWE-Playground pipeline in our experiments.

\label{sec:prompt}

\subsection{Prompt Template for Project Proposal}

\label{subsection:promptprojectproposal}

\begin{tcolorbox}[breakable, title=System Prompt for Project Proposal]
\small
We are building a framework to train SWE agents to solve software engineering problems.\\
The plan is to propose projects and tasks for the agents to solve, collect task-solving trajectories, and then use these trajectories to train the agents.\\

Right now, we are proposing multiple diverse projects for our agent to solve.\\
We would like to work on a diverse set of projects to train a generalist SWE-agent that is strong across a wide range of **complex computational tasks**.\\
Each task you propose should require multiple phases, 10+ distinct tasks, diverse technologies, and real-world complexity - think of projects that would take a professional developer 2-4 weeks to complete with multiple milestones and deliverables.\\

You should propose diverse Python projects that are **algorithmically complex and require substantial core logic implementation**. Projects should involve implementing sophisticated algorithms, data structures, computational methods, or complex system architectures from scratch.\\

**REQUIRED CHARACTERISTICS:**\\
- **Algorithm-heavy**: Must implement complex algorithms, data structures, or computational methods from scratch\\
- **Core logic implementation**: Cannot be solved by simply using existing libraries - requires substantial custom logic\\
- **Multi-component architecture**: Systems with multiple interacting components and complex state management\\
- **Non-trivial data processing**: Complex parsing, transformation, analysis, or generation of data\\
- **Custom protocols/formats**: Implementation of interpreters, compilers, parsers, or custom communication protocols\\

**COMPLEXITY REQUIREMENTS:**\\
- Projects should require implementing 200+ lines of core algorithmic logic (not just glue code)\\
- Must involve at least one of: advanced algorithms, mathematical computations, complex state machines, graph algorithms, optimization techniques, parsing/compilation, computer graphics, machine learning from scratch, or distributed algorithms\\
- Should require designing and implementing custom data structures or algorithms\\
- Cannot be completed by primarily calling existing library functions\\

**INTERFACE REQUIREMENT:**\\
- **Command-line interface only**: All projects must be operated via CLI with command-line arguments, flags, and text-based output\\
- **No GUI components**: Projects must not include graphical user interfaces, web interfaces, or visual components\\
- **Text-based interaction**: All input/output should be through files, command-line parameters, or terminal text\\
- **Scriptable and automatable**: CLI design should enable easy automation and testing\\

**LIBRARY CONSTRAINTS REQUIREMENT:**\\
Each project must include specific constraints on which libraries/frameworks cannot be used to ensure core algorithm implementation from scratch. These constraints should force developers to implement the fundamental algorithms rather than using existing solutions.\\

**CLEAR SPECIFICATION REQUIREMENT:**\\
Each project must be clearly defined with specific, concrete requirements that enable comprehensive test case development. This includes:\\
- **Specific input/output formats**: Clear definition of expected inputs and outputs\\
- **Concrete functionality**: Detailed description of what the system should do\\
- **Measurable performance criteria**: Specific performance requirements or benchmarks\\
- **Testable behavior**: Well-defined behaviors that can be verified through testing\\
- **Unambiguous success criteria**: Clear definition of what constitutes a successful implementation\\
- **Edge case handling**: Explicit requirements for error conditions and boundary cases\\

**AVOID:**\\
- Simple CRUD applications or basic data management systems\\
- Projects that are primarily UI-focused with minimal logic\\
- Basic file processing tools that use standard libraries\\
- Simple wrappers around existing libraries or APIs\\
- Standard desktop applications without algorithmic complexity\\
- Basic web applications with simple form handling\\
- Projects that can be completed with \textless 100 lines of core logic\\
- Vague or ambiguous project descriptions that don't specify concrete requirements\\
- Any projects requiring GUI, web interfaces, or visual components such as computer graphics related tasks\\
- Any projects requiring non-textual elements such as images or files as input and output\\
- Any projects aiming at some open-ended problems and implementation correctness cannot be easily evaluated\\

**GOOD PROJECT TYPES:**\\
- **Language processors**: Compilers, interpreters, code analyzers, DSL implementations\\
- **Complex data structures**: Database engines, distributed data structures\\
- **Mathematical tools**: Computer algebra systems, numerical computation libraries, statistical frameworks\\
- **Advanced parsers**: Complex format processors, protocol implementations, data transformation engines\\
- **Network protocols**: Custom communication protocols, distributed consensus algorithms\\
- **Optimization systems**: Constraint solvers, scheduling algorithms, resource allocation engines\\

Examples of appropriately complex project proposals with clear specifications:\\

\textless proposed\_project\textgreater Build a complete SQL database engine that supports a subset of SQL (SELECT, INSERT, UPDATE, DELETE, CREATE TABLE, CREATE INDEX) with B+ tree indexing, transaction management with ACID properties, and concurrent access. Must handle SQL parsing, query optimization with cost-based optimizer, join algorithms (nested loop, hash join), and support for integer, string, and float data types. The database should pass standard SQL compliance tests and handle concurrent transactions with proper isolation levels. CLI interface for SQL commands with options for database file, transaction mode, and performance statistics.\textless /proposed\_project\textgreater \\
\textless repo\_name\textgreater sql-database-engine\textless /repo\_name\textgreater \\
\textless programming\_language\textgreater Python\textless /programming\_language\textgreater \\
\textless constraints\textgreater Cannot use any existing database libraries (e.g., SQLite, PostgreSQL bindings). Must implement all SQL parsing, B+ tree structures, and transaction management from scratch without using ORM frameworks or query engines. Only basic file I/O operations allowed.\textless /constraints\textgreater \\

\textless proposed\_project\textgreater Create a compiler for a custom functional programming language with static typing, pattern matching, and recursion. Must implement lexical analysis, recursive descent parsing, type inference with Hindley-Milner algorithm, and bytecode generation. The language should support basic arithmetic, boolean operations, lists, tuples, functions, and algebraic data types. The compiler should detect and report syntax errors, type errors, and generate efficient bytecode that runs on a custom virtual machine. CLI interface with compilation flags, optimization levels, and detailed error reporting.\textless /proposed\_project\textgreater \\
\textless repo\_name\textgreater functional-language-compiler\textless /repo\_name\textgreater \\
\textless programming\_language\textgreater Python\textless /programming\_language\textgreater \\
\textless constraints\textgreater Cannot use any existing parser generators (e.g., PLY, ANTLR, Lark). Must implement all parsing, type inference, and code generation algorithms from scratch without compiler construction frameworks. Only basic string processing and file I/O allowed.\textless /constraints\textgreater \\
\end{tcolorbox}

\begin{tcolorbox}[breakable, title=User Prompt for Project Proposal]
\small
Please propose exactly \{\{ num\_projects \}\} diverse projects for the agent to work on, along with names for each project repo.\\
Format your output as a list of exactly \{\{ num\_projects \}\} projects, each in the following format:\\

Project 1:\\
\textless proposed\_project\textgreater ...\textless /proposed\_project\textgreater \\
\textless repo\_name\textgreater ...\textless /repo\_name\textgreater \\
\textless programming\_language\textgreater ...\textless /programming\_language\textgreater \\
\textless constraints\textgreater ...\textless /constraints\textgreater \\

Project 2:\\
\textless proposed\_project\textgreater ...\textless /proposed\_project\textgreater \\
\textless repo\_name\textgreater ...\textless /repo\_name\textgreater \\
\textless programming\_language\textgreater ...\textless /programming\_language\textgreater \\
\textless constraints\textgreater ...\textless /constraints\textgreater \\

... (continue for all projects)\\

Each proposed project should be **algorithmically complex and computationally challenging** with the following characteristics:\\
- **Algorithm-heavy**: Must implement complex algorithms, data structures, or computational methods from scratch\\
- **Core logic implementation**: Cannot be solved by simply using existing libraries - requires substantial custom logic\\
- **Multi-component architecture**: Systems with multiple interacting components and complex state management\\
- **Non-trivial data processing**: Complex parsing, transformation, analysis, or generation of data\\
- **Custom protocols/formats**: Implementation of interpreters, compilers, parsers, or custom communication protocols\\

**INTERFACE REQUIREMENT:**\\
- **Command-line interface only**: All projects must be operated via CLI with command-line arguments, flags, and text-based output\\
- **No GUI components**: Projects must not include graphical user interfaces, web interfaces, or visual components\\
- **Text-based interaction**: All input/output should be through files, command-line parameters, or terminal text\\
- **Scriptable and automatable**: CLI design should enable easy automation and testing\\

**LIBRARY CONSTRAINTS REQUIREMENT:**\\
Each project must include specific constraints in the \textless constraints\textgreater  section about which libraries/frameworks cannot be used. These constraints should force developers to implement fundamental algorithms from scratch rather than using existing solutions. Examples:\\
- Computer algebra systems: "Cannot use numpy, scipy, or sympy"\\
- Machine learning: "Cannot use TensorFlow, PyTorch, or scikit-learn"\\
- Cryptography: "Cannot use existing crypto libraries like cryptography or pycrypto"\\
- Database engines: "Cannot use SQLite, PostgreSQL, or any existing database libraries"\\

**CLEAR SPECIFICATION REQUIREMENT:**\\
Each project must be clearly defined with specific, concrete requirements that enable comprehensive test case development:\\
- **Specific input/output formats**: Clearly define expected inputs and outputs with exact formats\\
- **Concrete functionality**: Detailed description of what the system should do with specific behaviors\\
- **Measurable performance criteria**: Include specific performance requirements, benchmarks, or constraints\\
- **Testable behavior**: Well-defined behaviors that can be verified through unit tests and integration tests\\
- **Unambiguous success criteria**: Clear definition of what constitutes a successful implementation\\
- **Edge case handling**: Explicit requirements for error conditions, boundary cases, and failure modes\\

All projects should use Python as the programming language. Projects should focus on implementing sophisticated computational systems with well-defined, testable specifications accessible only through CLI.\\

**DIVERSITY REQUIREMENT:**\\
Ensure maximum diversity across the proposed projects by:\\
- **Distributing across complexity areas**: Include projects from each of the complexity areas listed above, not clustering around just a few areas\\
- **Varying system types**: Mix low-level systems, high-level algorithms, and mid-level tools\\
- **Different problem domains**: Include projects spanning various domains\\
- **Balanced complexity focus**: Some projects should focus on data structures, others on algorithms, others on system design, others on mathematical computation\\
- **Avoid repetition**: Each project should address distinctly different computational challenges and use cases\\

**AVOID** (these are too simple, too vague, or inappropriate):\\
- Basic CRUD applications or simple data management\\
- Standard desktop/web apps without algorithmic complexity  \\\\
- File processing tools using standard libraries\\
- Simple wrappers around existing libraries\\
- Projects primarily focused on UI with minimal logic\\
- Basic automation scripts or simple utilities\\
- Vague or ambiguous project descriptions that don't specify concrete requirements\\
- Any projects requiring GUI, web interfaces, or visual components such as computer graphics related tasks\\
- Any projects requiring non-textual elements such as images or files as input and output\\
- Any projects aiming at some open-ended problems and implementation correctness cannot be easily evaluated\\

Generate the projects as you are required. You are encouraged to be more creative and propose some projects that are not among the recommendation types.\\
\end{tcolorbox}

\subsection{Prompt Template for Task Proposal}
\label{subsection:prompttaskproposal}

\begin{tcolorbox}[breakable, title=System Prompt for Task Proposal]
\small
We are building a framework to train SWE agents to solve software engineering problems.\\
The plan is to propose projects and tasks for the agents to solve, collect task-solving trajectories, and then use these trajectories to train the agents.\\

We are currently in the middle of the proposal stage. Particularly, a project proposer has already proposed a project idea.\\
Now, it is your job to do the following:\\

Propose a sequence of tasks to solve for this project. The proposed tasks should form a closed loop, meaning it could be self-contained without requiring any external dependencies, such as calls to external APIs or services. All necessary components and resources should be included within the project itself to ensure full autonomy and reproducibility.\\

Create a comprehensive task instruction document following this exact format:\\

\# Project Description\\

[Description of the project]\\

\# Project Instruction\\

[Instruction to finish the Project]\\

\# Detailed Documentation\\

\#\# Phase X: [Phase Name]\\

Organize your tasks into logical phases (e.g., Phase 1: Foundation, Phase 2: Core Features, Phase 3: Advanced Features. Each phase should have:\\

**Goal:** Clear description of what this phase accomplishes.\\

\#\#\# Module X.Y: [Module name]\\

Within each phase, group related tasks into modules. For each task, provide:\\

\#\#\#\# Task X.Y.Z: [Task Name]\\

- **Description:** Clear, detailed description of what needs to be implemented\\
- **Dependencies:** List of prerequisite tasks (use task numbers like 1.1.1, 1.2.3, etc.)\\
- **Difficulty:** Rating from 1/5 to 5/5\\
- **Unit Tests:**\\
  - **Code Tests:** (when applicable)\\
    - **TestName:** Specific unit test descriptions that verify functionality\\
  - **Visual Tests:** (when applicable for UI/visual components)
    - **TestName:** Integration test descriptions for visual verification\\

Each task should be concrete and at the granularity of a single pull request. Tasks should build upon each other logically, with clear dependency chains.\\

Make sure that the total number of phases should not exceed five. Focus on implementation tasks only - do not include documentation, deployment, evaluation, performance optimization (unless the project itself requires) etc. into the tasks. All phases should be dedicated to building the core functionality of the project.\\

Please follow the format of the exactly and avoid any extra information, as this file will be automatically parsed later in our pipeline. Remember to have all Phase, Module and Task present.\\
\end{tcolorbox}

\begin{tcolorbox}[breakable, title=User Prompt for Task Proposal]
\small
Please propose the tasks for the project: \{\{ project\_description \}\}.\\
Notice here are some constraints for the project: \{\{ constraints \}\}
Format your output as:\\

\textless tasks\textgreater 
\# Project Description\\

[Description of the project]\\

\# Project Instruction\\

[Instruction to finish the project]\\

\# Detailed Documentation\\

\#\# Phase 1: [Phase Name]\\

**Goal:** [Description of what this phase accomplishes]\\

\#\#\# Module 1.1: [Module Name]\\

\#\#\#\# Task 1.1.1: [Task Title]\\

- **Description:** [Detailed description of what needs to be implemented]\\
- **Dependencies:** [List prerequisite task numbers, e.g., None, 1.1.1, 1.2.3]\\
- **Difficulty:** [Rating from 1/5 to 5/5]\\
- **Unit Tests:**\\
  - **Code Tests:**\\
    - **[TestName]:** [First code test that verifies functionality]\\
    - **[TestName]:** [Second code test that verifies functionality]\\
  - **Visual Tests:**\\
    - **[TestName]:** [First visual test that verifies functionality]\\

\#\#\#\# Task 1.1.2: [Task Title]\\

- **Description:** [Detailed description of what needs to be implemented]\\
- **Dependencies:** [List prerequisite task numbers]\\
- **Difficulty:** [Rating from 1/5 to 5/5]\\
- **Unit Tests:**\\
  - **Code Tests:**\\
    - **[TestName]:** [First code test that verifies functionality]\\

\#\#\# Module 1.2: [Module Name]\\

\#\#\#\# Task 1.2.1: [Task Title]\\

- **Description:** [Detailed description of what needs to be implemented]\\
- **Dependencies:** [List prerequisite task numbers]\\
- **Difficulty:** [Rating from 1/5 to 5/5]\\
- **Unit Tests:**\\
  - **Visual Tests:**\\
    - **[TestName]:** [First visual test that verifies functionality]\\

\#\# Phase 2: [Phase Name]\\

**Goal:** [Description of what this phase accomplishes]\\

\#\#\# Module 2.1: [Module Name]\\

\#\#\#\# Task 2.1.1: [Task Title]\\

- **Description:** [Detailed description of what needs to be implemented]\\
- **Dependencies:** [List prerequisite task numbers from previous phases]\\
- **Difficulty:** [Rating from 1/5 to 5/5]\\
- **Unit Tests:**\\
  - **Code Tests:** N/A\\

\#\# Phase 3: [Phase Name]\\

**Goal:** [Description of what this phase accomplishes]\\

\#\#\# Module 3.1: [Module Name]\\

\#\#\#\# Task 3.1.1: [Task Title]\\

- **Description:** [Detailed description of what needs to be implemented]\\
- **Dependencies:** [List prerequisite task numbers from previous phases]\\
- **Difficulty:** [Rating from 1/5 to 5/5]\\
- **Unit Tests:**\\
  - **Code Tests:**\\
    - **[TestName]:** [First code test that verifies functionality]
\textless /tasks\textgreater \\

Make sure that your output is wrapped in \textless tasks\textgreater  and \textless /tasks\textgreater  tags.\\
\end{tcolorbox}

In our experiments during task proposal, model output is sometimes truncated without having closing tags. We thus use the following prompt to continue with the unfinished generation.

\begin{tcolorbox}[breakable, title=User Prompt for Task Proposal (Continue)]
\small
You are generating tasks for the following project:\\

**Project Description:** \{\{ project\_description \}\}\\
**Project Constraints:** \{\{ constraints \}\}\\

Your previous response was:\\
\textasciigrave \textasciigrave \textasciigrave \\
\{\{ response \}\}\\
\textasciigrave \textasciigrave \textasciigrave \\

Your response appears to be incomplete as it doesn't have a closing \textasciigrave \textless /tasks\textgreater \textasciigrave  tag. Please continue generating tasks from exactly where you left off. Do not repeat any content from your previous response - just continue with the next task(s) and make sure to properly close the \textasciigrave \textless /tasks\textgreater \textasciigrave  tag when you're finished.\\

Continue from where you left off:\\
\end{tcolorbox}

Also in this stage, we generate a detailed unit test checklist for each proposed task, and the used prompt is provided below:

\begin{tcolorbox}[breakable, title=System Prompt for Checklist Generation]
\small
We are building a framework to train SWE agents to solve software engineering problems.\\
The plan is to propose projects and tasks for the agents to solve, collect task-solving trajectories, and then use these trajectories to train the agents.\\

We are currently in the final stage. Particularly, a project and all the tasks have been proposed.\\
Now, it is your job to do propose unit tests based on the given task requirements.\\
\end{tcolorbox}

\begin{tcolorbox}[breakable, title=User Prompt for Checklist Generation]
\small
Project Description: \{\{ project\_description \}\}\\

Tasks for the project:\\

\{\{ tasks\_prompt \}\}\\

All of the proposed unit tests:\\

\{\{ previous\_unit\_tests \}\}\\

Now you should propose the detailed unit tests for:\\
\{\{ unit\_test\_prompt \}\}\\

You only need to propose the unit tests and do not need to implement them. You should describe:\\

1. **Test Case Name and Purpose**: A clear, descriptive name for each test case and what specific functionality it verifies.\\

2. **Test Scenarios**: Cover different scenarios including:\\
   - Normal/happy path cases\\
   - Edge cases and boundary conditions\\
   - Error handling and exception cases\\

3. **Assertions**: Specify what should be verified/asserted in each test case.\\

Also, please also take previous unit tests into consideration. To be specific:\\

1. **Avoid Duplication**: Ensure the new unit tests don't duplicate existing test cases. Review the previous tests to understand what has already been covered.\\

2. **Consider Dependencies**: Be aware of how the new functionality interacts with previously tested components and ensure integration points are properly tested.\\

3. **Ensure No Conflicts**: Make sure the new tests won't interfere with or break existing test cases.\\

Please provide comprehensive test coverage that ensures the reliability and correctness of the implementation. Be precise with your output.\\
\end{tcolorbox}

\subsection{Prompt Template for Repository Setup}

\begin{tcolorbox}[breakable, title=OpenHands Prompt for Repository Setup]
\small
We are building a framework to train SWE agents to solve software engineering problems.\\
The plan is to propose projects and tasks for the agents to solve, collect task-solving trajectories, and then use these trajectories to train the agents.\\

We are currently in the middle of the proposal stage. Particularly, a project proposer has already proposed a project idea and the tasks for this project has also been proposed in \textasciigrave tasks.md\textasciigrave .\\

**Project Description:** \{\{ project\_description \}\}\\

Notice here are some constraints for the project: \{\{ constraints \}\}\\

Now, it is your job to do the following:\\

1. Set up the basic repository structure for the project.\\

Create the initial project structure with the following components:\\

- **Project Configuration Files**: Set up appropriate configuration files based on the project type:\\
  - For Python projects: requirements.txt, setup.py, or pyproject.toml\\
  - For C++ projects: CMakeLists.txt, Makefile, or similar\\
  - For Rust projects: Cargo.toml\\

- **Directory Structure**: Create the basic folder structure:\\
  - \textasciigrave src/\textasciigrave  or \textasciigrave app/\textasciigrave  for source code\\
  - \textasciigrave tests/\textasciigrave  for test files\\
  - \textasciigrave docs/\textasciigrave  for documentation\\
  - \textasciigrave assets/\textasciigrave  for static resources (images, data files, etc.)\\

- **Initial Source Files**: Create placeholder files to establish the project structure:\\
  - Main entry point (main.py, main.cpp, etc.)\\
  - Basic module structure\\
  - Initial test files\\

The repository should be immediately buildable and runnable, but you should not implement any anythings involved as tasks in \textasciigrave task.md\textasciigrave , which means your implemented setup must not pass any unit test.\\

You may create some directories and files, but do not write substantial content into them. The setup should only include the minimal structure needed to support the implementation of tasks.\\

Also, unit tests should not be setup in this stage.\\

Notice that if you would like to initialize some classes and their functions, you should raise a \textasciigrave NotImplementedError\textasciigrave  in each method body and do not provide any actual implementation, instead of simply leave a ``pass''.\\

2. Prepare Docker setup for the project.\\

- Modify \textasciigrave Dockerfile\textasciigrave  and any other necessary files (such as \textasciigrave .dockerignore\textasciigrave ) to enable containerized development of the project.\\
- Ensure the Docker setup installs all required build tools and dependencies for the project to compile and run, and most importantly for further development.\\
- The Docker container must be persistent for further development. Do not modify \textasciigrave CMD [``tail'',``-f'',``/dev/nul'']\textasciigrave  is Dockerfile.\\
- For Python setup, you should create a new conda environment instead of directly installing all the packages. Therefore, use miniconda and environment.yml for setting up.\\

A template Dockerfile tailored to the project's programming language has been provided in the repository. Please review and update this Dockerfile as needed to ensure it fully supports the specific requirements and dependencies of your project.\\
\end{tcolorbox}

\subsection{Prompt Template for Unit Test Generation}

\begin{tcolorbox}[breakable, title=OpenHands Prompt for Unit Test Generation]
\small
We are building a framework to train SWE agents to solve software engineering problems.\\
The plan is to propose projects and tasks for the agents to solve, collect task-solving trajectories, and then use these trajectories to train the agents.\\

We are currently in the final of the proposal stage. Particularly, a project proposer has already proposed a project idea to tackle, and the documentation for the project has already been set up.\\

**Task:** \{\{ project\_task \}\}\\

Now, it is your job to create unit tests for the project based on the documentation:\\

For each Task X.Y.Z, a list of unit tests are contained in the documentation. Unit tests are put into two categories: Code Tests and Visual Tests.\\

For Code Tests, you need to write the actual unit test code that verifies the functionality described in the test description. These should be proper unit tests with assertions that check specific behaviors, inputs, and outputs.\\

For Visual Tests, you need to create simple test programs that call multimodal agents to provide visual verification of UI components, rendering outputs, and user interface behaviors that cannot be easily tested through code assertions alone.\\

Please create unit test files for each task that has unit tests defined in the documentation. Organize the tests logically and ensure they cover all the test cases mentioned in the task descriptions.\\

The unit tests should be written in the appropriate language for the project and follow standard testing conventions and frameworks.\\

Along with the unit tests functions you have written, a bash script \textasciigrave tests/X.Y.Z.sh\textasciigrave  is needed. All unit tests will be finally executed via this bash script. Each bash script should:\\

1. Set up the necessary environment (compile the project if needed)\\
2. Run the specific unit tests for that task\\
3. Handle both code tests and visual tests appropriately\\
4. Provide clear output indicating test results (pass/fail)\\
5. Exit with appropriate status codes (0 for success, non-zero for failure)\\

For visual tests, the bash script should run the programs that invoke a multimodal agent to perform visual verification of the test outputs.\\

The scripts will be executed under root directory via \textasciigrave bash ./tests/1.1.1.sh\textasciigrave , and this bash script should run \textasciigrave ./tests/test\_1\_1\_1.py\textasciigrave  via pytest, which implements the unit test logic. So you should ensure that the scripts can be run in this way.\\

You do not need to install packages for unit tests. Also, if the tests run into cases such as Import Error or Not Implemented Error, you should treat it as a failure case instead of skipping it.\\

All Tasks should be covered by unit tests except those marked with N/A. You do not need to care about whether your unit tests need to import the implemented modules. You just need to be responsible for implementing unit tests, and the implementation of functions will be based on your unit tests. You shall not leave TODOs in unit tests.\\

We are generating unit tests task by task, and currently you need to implement:\\

\{\{ unit\_test\_prompt \}\}\\

Only work on generating code for this unit test and do not care about others and no documentation for unit tests is needed. You MUST NOT implement code except for unit tests.\\

Please note that your task is to generate unit tests that can verify the correctness of the code that others will implement. You should not make your unit tests pass with the current code implementation.
You should also NEVER expect the raise of NotImplementedError in unit tests as this will encourage the behavior of leaving functions unfinished.\\

You should ensure the tests you implement correctly verify all the functions for different projects:\\
- **Mathematical tools**: Unit tests should ensure numerical accuracy, handle edge cases like division by zero, infinity, and NaN values, test mathematical properties and invariants, and validate algorithm correctness across different input ranges\\
- **Language processors**: Unit tests should verify parsing accuracy, tokenization correctness, syntax validation, error handling for malformed input, and proper handling of different language constructs and edge cases\\
- **Complex data structures**: Database engines, distributed data structures should be tested for data integrity, concurrent access safety, performance characteristics, memory management, and correctness of complex operations like joins, transactions, and consistency guarantees\\
- **Advanced parsers**: Unit tests should validate parsing of complex grammars, error recovery mechanisms, abstract syntax tree generation, semantic analysis correctness, and handling of ambiguous or malformed input\\
- **Network protocols**: Unit tests should verify message serialization/deserialization, protocol state management, error handling for network failures, timeout behaviors, and compliance with protocol specifications\\
- **Optimization systems**: Unit tests should test convergence properties, objective function evaluation, constraint satisfaction, performance benchmarks, and correctness of optimization algorithms across various problem instances\\
Make sure that the functionality is thoroughly tested, rather than simply verifying that the output format matches expectations.\\
\end{tcolorbox}

\subsection{Prompt Template for Functionality Implementation}

\begin{tcolorbox}[breakable, title=Prompt for Functionality Implementation Generation]
\small
You are now given a project directory and required to finish the project task by task.\\
You have finished all tasks before \{\{ task\_number \}\}. Now it's your task to finish Task \{\{ task\_number \}\}.\\

\textless task\_description\textgreater \\
\{\{ task\_description \}\}\\
\textless /task\_description\textgreater \\

You should finish the task based on the given \textless task\_description\textgreater . Also, here are some constraints that you should follow:\\
\{\{ constraints \}\}\\

In the given repo, there are unit tests under \textasciigrave /tests\textasciigrave , and you can run via \textasciigrave bash tests/\{\{ task\_number \}\}.sh\textasciigrave .\\
However, you should NEVER first read and run the unit tests when you start with the task.\\
First directly read the source code and finish with the implementation without accessing the unit tests. During this process, you can write some your own test code or scripts to verify your implementation, but DO NOT read the content of the given tests.\\
After you think you have finished your implementation, you can run and read the provided unit tests to further verify and also correct the issues of your implementation.\\

Passing all the tests of \{\{ task\_number \}\} and also all tests for previous tasks means you have finished this task.\\
You MUST NOT modify any content under \textasciigrave /tests\textasciigrave .\\
You also do not need to proceed to next tasks.\\
\end{tcolorbox}

\subsection{Capability Specific Adaption}
\label{subsection:promptspecific}

\begin{tcolorbox}[breakable, title=System Prompt for Issue Proposal]
\small
We are building a framework to train SWE agents to solve software engineering problems.\\
The plan is to propose projects and tasks for the agents to solve, collect task-solving trajectories, and then use these trajectories to train the agents.\\

You are tasked with creating realistic software bugs/issues that will be injected into working code to generate training data for SWE agents. The current codebase is fully functional and passes all unit tests, but we need to introduce specific issues that agents can learn to fix.\\

**Your Role:** Software Quality Assurance Engineer identifying realistic bugs\\

**Context:** You have access to:\\
- Complete unit test suite that validates the functionality\\
- Task documentation describing the intended behavior\\

**Task:** Generate a realistic, specific bug description that:\\

**CRITICAL REQUIREMENTS:**\\
1. **Realistic \& Common:** The issue should represent bugs commonly found in real software development (off-by-one errors, null pointer exceptions, incorrect conditionals, edge case handling, etc.)\\

2. **Test-Breaking:** Must cause at least one unit test to fail, but preferably multiple related tests\\

3. **Implementable:** Should be clear enough that another developer could introduce this exact bug by making specific, targeted code changes\\

4. **Scoped Appropriately:** \\
   - Focus on logical errors, not syntax errors\\
   - Affect specific functionality without breaking the entire system\\
   - Target edge cases, boundary conditions, or specific input scenarios\\

5. **Detailed \& Actionable:** Include:\\
   - Specific conditions under which the bug occurs\\
   - Expected vs actual behavior\\
   - Concrete examples of problematic inputs/scenarios\\
   - Clear description of what's going wrong\\

**Issue Types to Consider:**\\
- Off-by-one errors in loops or array indexing\\
- Incorrect conditional logic (wrong operators, missing conditions)\\
- Edge case handling failures (empty inputs, boundary values)\\
- Type conversion or data validation issues\\
- Resource management problems\\
- Concurrency issues (if applicable)\\
- Algorithm implementation errors\\

**Avoid:**\\
- Syntax errors or compilation issues\\
- Massive architectural changes\\
- Issues requiring external dependencies\\
- Vague descriptions like ``doesn't work''\\
\end{tcolorbox}

\begin{tcolorbox}[breakable, title=User Prompt for Issue Proposal]
\small
**PROJECT:** \{\{ project\_description \}\}\\

**TASK DOCUMENTATION:**\\
\{\{ test\_prompt \}\}\\

**UNIT TEST CODE:**\\
\textasciigrave \textasciigrave \textasciigrave\\
\{\{ test\_code \}\}\\
\textasciigrave \textasciigrave \textasciigrave\\

**INSTRUCTIONS:**\\
Analyze the provided unit tests and task documentation to understand what the code is supposed to do. Then propose a realistic bug that would cause some of these tests to fail.\\

**Your Analysis Process:**\\
1. **Understand the Requirements:** What functionality is being tested?\\
2. **Identify Test Scenarios:** What specific cases are the unit tests checking?\\
3. **Find Vulnerable Points:** Where could realistic bugs be introduced?\\
4. **Craft the Bug:** Describe a specific, implementable issue\\

**Issue Description Requirements:**\\
- **Be Specific:** Include exact conditions, inputs, or scenarios where the bug occurs\
- **Be Realistic:** Choose common bug patterns that real developers make\\\
- **Be Testable:** Ensure the bug would cause specific unit tests to fail\\
- **Be Implementable:** Provide enough detail that a developer could introduce this exact bug\\

**Good Example Format:**\\
``The function incorrectly handles empty string inputs. When an empty string is passed as input, the function should return an empty list according to the specification, but instead it returns None. This occurs because the validation logic at the beginning of the function has an incorrect condition that treats empty strings as invalid input.''\\

**Your Task:**\\
Generate TWO things:\\
1. **Issue**: A technical description for applying the bug to code\\
2. **Description**: A user-friendly bug report like a GitHub issue or pull request\\

**Format Requirements:**\\

\textless issue\textgreater \\
\lbrack Technical description for applying the bug to code. Be specific about what code needs to be changed, what conditions trigger the bug, and what should happen vs what actually happens. This will be used to inject the bug into the codebase.\rbrack \\
\textless /issue\textgreater \\

\textless description\textgreater \\
\lbrack User-friendly bug report like a GitHub issue. Include a clear title-like summary, description of the problem, and any relevant context. Write this as if a real developer discovered the bug and is reporting it. The description should not be too detailed. For example, you should not provide the code to reproduce the issue\rbrack \\
\textless /description\textgreater \\

**Example Format:**\\
\textless issue\textgreater \\
The validation function incorrectly handles empty string inputs. When an empty string is passed to validate\_input(), it should return False according to the specification, but the current condition checks 'if not input' which treats empty strings as invalid when they should be valid for this specific case.\\
\textless /issue\textgreater \\

\textless description\textgreater \\
Cannot validate empty strings in input validator\\

I'm having trouble with the input validation function. When I pass an empty string as input, it returns \textasciigrave False\textasciigrave , but according to the documentation and expected behavior, empty strings should be considered valid input for this validator.\\

This worked in previous versions but seems to be broken now. The function appears to be treating empty strings the same as null/undefined values, which is incorrect according to the specification.\\
\textless /description\textgreater \\
\end{tcolorbox}

\begin{tcolorbox}[breakable, title=OpenHands Prompt for SWE-bench Issue Application]
\small
You are given a project directory with working code and a specific issue that needs to be applied to the codebase.\\

Your task is to read the issue description and apply the described problem to the current working code. This means you should INTRODUCE the bug or issue described, causing the existing unit tests to fail.\\

**Issue to Apply:**\\
\{\{ issue\_description \}\}\\

**Project Description:**\\
\{\{ project\_description \}\}\\

**Task Information:**\\
Task \{\{ task\_number \}\}: \{\{ task\_description \}\}\\

**Instructions:**
1. First, understand the current working implementation by reading the relevant source code\\
2. Identify where the issue should be applied based on the issue description\\
3. Make the necessary code changes to introduce the described issue/bug\\
4. The goal is to make the unit tests fail by introducing this specific issue\\
5. Run the unit tests of this task to ensure they no longer pass. DO NOT edit unit test files.\\
6. Keep changes minimal and focused on the specific issue described\\
7. Ensure the code still compiles/runs but with the intended bug/issue\\

**Important:** You are introducing a problem into working code, not fixing it. The unit tests should fail after you apply this issue.\\
\end{tcolorbox}

\begin{tcolorbox}[breakable, title=OpenHands Prompt for SWT-Bench Issue Application]
\small
You are given a project directory with working code and a specific issue that needs to be applied to the codebase.\\

Your task is to read the issue description and apply the described problem to the current working code. Also, you should modify the test files so that the issue cannot be detected by the tests. This means you should INTRODUCE the bug or issue described, but still making that all the unit tests under `tests/` can pass by modifying the test scripts to remove the specific test functions related to this issue and also modify other functions is this issue will also cause them to fail.\\

**Issue to Apply:**\\
\{\{ issue\_description \}\}\\

**Project Description:**\\
\{\{ project\_description \}\}\\

**Task Information:**\\
Task \{\{ task\_number \}\}: \{\{ task\_description \}\}\\

**Instructions:**\\
1. First, understand the current working implementation by reading the relevant source code\\
2. Identify where the issue should be applied based on the issue description\\
3. Make the necessary code changes to introduce the described issue/bug\\
4. The goal is to make the unit tests fail by introducing this specific issue\\
5. Run the unit tests of this task to ensure they no longer pass. DO NOT edit unit test files.\\
6. Keep changes minimal and focused on the specific issue described\\
7. Ensure the code still compiles/runs but with the intended bug/issue\\

**Important:** You are introducing a problem into working code, not fixing it. Remember that you should remove the related test functions and keep test scripts pass.\\
\end{tcolorbox}

\section{Case Study for Generated Tasks}

\label{sec:casestudy}

\subsection{Case Study for Project Proposal}

\begin{tcolorbox}[breakable, title=Project Proposal for Project ``version-control-system'']
\small
\textbf{Repo Name:}\\
version-control-system\\

\textbf{Programming Language:}\\
Python\\

\textbf{Project Description:}\\
Create a Git-like version control system with support for repositories, commits, branches, merging, and diff algorithms. Must implement object storage with SHA-1 hashing, three-way merge algorithm, conflict resolution, branch management, and repository compression. The system should handle binary files, large repositories (>10MB), and maintain complete history integrity. CLI interface with commands for init, add, commit, branch, merge, log, diff, and status operations matching Git's interface.\\

\textbf{Constraints:}\\
Cannot use existing VCS libraries (e.g., GitPython, dulwich). Must implement all version control algorithms, diff algorithms, and merge strategies from scratch without using version control frameworks. Only basic file system operations and hashing allowed.\\
\end{tcolorbox}

\begin{tcolorbox}[breakable, title=Project Proposal for Project ``filesystem-simulator'']
\small
\textbf{Repo Name:}\\
filesystem-simulator\\

\textbf{Programming Language:}\\
Python\\

\textbf{Project Description:}\\
Create a file system simulator supporting ext4-like features including inodes, directory structures, file allocation, journaling, and defragmentation. Must implement block allocation algorithms, metadata management, crash recovery, and file system consistency checks. The simulator should handle file systems up to 1GB size and support concurrent file operations. CLI interface for mounting/unmounting, file operations (create, read, write, delete), directory navigation, and file system maintenance commands.\\

\textbf{Constraints:}\\
Cannot use existing file system libraries or implementations. Must implement all file system algorithms, block management, and metadata structures from scratch without using file system frameworks. Only basic file I/O for the underlying storage file allowed.\\
\end{tcolorbox}

\subsection{Case Study for Task Proposal}

\begin{tcolorbox}[breakable, title=Task Proposal for Project ``filesystem-simulator'']
\small
\# Project Description \\

A comprehensive computer algebra system (CAS) that implements symbolic mathematics, polynomial arithmetic, and calculus operations. The system supports expression parsing from LaTeX format, algebraic simplification, differentiation, integration, and equation solving. It includes matrix operations, linear algebra capabilities, and numerical methods for root finding. The system can solve systems of equations with 10+ variables and integrate complex functions symbolically, providing step-by-step solutions and numerical approximations. \\

\# Project Instruction \\

Build a complete computer algebra system from scratch that can parse LaTeX mathematical expressions, perform symbolic computations, and provide detailed step-by-step solutions. The system should handle polynomial arithmetic, calculus operations, matrix computations, and solve complex mathematical problems including large systems of equations and symbolic integration. \\

\# Detailed Documentation \\

\#\# Phase 1: Foundation and Expression System \\

**Goal:** Establish the core expression representation system and basic parsing capabilities to handle mathematical expressions. \\

\#\#\# Module 1.1: Core Expression Framework \\

\#\#\#\# Task 1.1.1: Expression Tree Data Structure \\

- **Description:** Implement a hierarchical expression tree structure to represent mathematical expressions with nodes for variables, constants, operators, and functions. Include support for expression metadata and type information. \\
- **Dependencies:** None \\
- **Difficulty:** 3/5 \\
- **Unit Tests:** \\
  - **Code Tests:** \\
    - **ExpressionNodeCreation:** Verify creation of different node types (variable, constant, operator, function) \\
    - **ExpressionTreeTraversal:** Test tree traversal methods (preorder, postorder, inorder) \\
    - **ExpressionEquality:** Test expression equality comparison and hashing \\

\#\#\#\# Task 1.1.2: Basic Expression Operations \\

- **Description:** Implement fundamental operations on expressions including copying, substitution, and basic structural manipulations. Add support for expression validation and type checking. \\
- **Dependencies:** 1.1.1 \\
- **Difficulty:** 2/5 \\
- **Unit Tests:** \\
  - **Code Tests:** \\
    - **ExpressionCopy:** Test deep copying of expression trees \\
    - **ExpressionSubstitution:** Verify variable substitution functionality \\
    - **ExpressionValidation:** Test expression structure validation \\

\#\#\# Module 1.2: LaTeX Parser \\

\#\#\#\# Task 1.2.1: LaTeX Tokenizer \\

- **Description:** Create a tokenizer that converts LaTeX mathematical expressions into tokens, handling mathematical symbols, operators, functions, and special LaTeX commands. Support common LaTeX mathematical notation including fractions, powers, and Greek letters. \\
- **Dependencies:** 1.1.1 \\
- **Difficulty:** 3/5 \\
- **Unit Tests:** \\
  - **Code Tests:** \\
    - **BasicTokenization:** Test tokenization of simple expressions \\
    - **ComplexTokenization:** Test tokenization of complex LaTeX with fractions and powers \\
    - **ErrorHandling:** Test handling of malformed LaTeX input \\

\#\#\#\# Task 1.2.2: LaTeX Parser Engine \\

- **Description:** Implement a recursive descent parser that converts tokenized LaTeX into expression trees. Handle operator precedence, parentheses, and LaTeX-specific constructs like  \verb|\|frac,  \verb|\|sqrt, and  \verb|\|sum. \\
- **Dependencies:** 1.2.1 \\
- **Difficulty:** 4/5 \\
- **Unit Tests:** \\
  - **Code Tests:** \\
    - **BasicParsing:** Test parsing of arithmetic expressions \\
    - **FractionParsing:** Test parsing of LaTeX fractions \\
    - **FunctionParsing:** Test parsing of mathematical functions \\

\#\#\# Module 1.3: Expression Display System \\

\#\#\#\# Task 1.3.1: Expression Renderer \\

- **Description:** Create a system to render expressions back to LaTeX format and human-readable text format. Support proper formatting with parentheses, operator precedence, and mathematical notation. \\
- **Dependencies:** 1.1.2 \\
- **Difficulty:** 2/5 \\
- **Unit Tests:** \\
  - **Code Tests:** \\
    - **LaTeXRendering:** Test conversion of expressions back to LaTeX \\
    - **TextRendering:** Test conversion to human-readable text format \\
    - **ParenthesesHandling:** Test proper parentheses placement \\

\#\# Phase 2: Algebraic Operations and Simplification \\

**Goal:** Implement core algebraic operations including polynomial arithmetic and expression simplification algorithms. \\

\#\#\# Module 2.1: Polynomial System \\

\#\#\#\# Task 2.1.1: Polynomial Representation \\

- **Description:** Implement a specialized polynomial class with support for multivariate polynomials, coefficient management, and degree calculations. Include efficient storage and manipulation of polynomial terms. \\
- **Dependencies:** 1.1.2 \\
- **Difficulty:** 3/5 \\
- **Unit Tests:** \\
  - **Code Tests:** \\
    - **PolynomialCreation:** Test creation of univariate and multivariate polynomials \\
    - **DegreeCalculation:** Test degree calculation for various polynomial types \\
    - **CoefficientAccess:** Test coefficient retrieval and modification \\

\#\#\#\# Task 2.1.2: Polynomial Arithmetic \\

- **Description:** Implement polynomial addition, subtraction, multiplication, and division operations. Include support for polynomial GCD calculation and factorization algorithms. \\
- **Dependencies:** 2.1.1 \\
- **Difficulty:** 4/5 \\
- **Unit Tests:** \\
  - **Code Tests:** \\
    - **BasicArithmetic:** Test addition, subtraction, multiplication of polynomials \\
    - **PolynomialDivision:** Test polynomial long division \\
    - **GCDCalculation:** Test greatest common divisor calculation \\

\#\#\# Module 2.2: Algebraic Simplification \\

\#\#\#\# Task 2.2.1: Basic Simplification Rules \\

- **Description:** Implement fundamental algebraic simplification rules including constant folding, identity operations, and basic algebraic identities. Create a rule-based simplification engine. \\
- **Dependencies:** 1.1.2 \\
- **Difficulty:** 3/5 \\
- **Unit Tests:** \\
  - **Code Tests:** \\
    - **ConstantFolding:** Test simplification of constant expressions \\
    - **IdentitySimplification:** Test simplification using algebraic identities \\
    - **ZeroOneSimplification:** Test simplification involving zero and one \\

\#\#\#\# Task 2.2.2: Advanced Simplification \\

- **Description:** Implement advanced simplification techniques including trigonometric identities, logarithmic simplification, and rational expression simplification. Add support for pattern matching and rewriting. \\
- **Dependencies:** 2.2.1 \\
- **Difficulty:** 4/5 \\
- **Unit Tests:** \\
  - **Code Tests:** \\
    - **TrigonometricSimplification:** Test simplification of trigonometric expressions \\
    - **LogarithmicSimplification:** Test simplification of logarithmic expressions \\
    - **RationalSimplification:** Test simplification of rational expressions \\

\#\# Phase 3: Calculus Operations \\

**Goal:** Implement differentiation and integration capabilities with support for symbolic computation and step-by-step solutions. \\

\#\#\# Module 3.1: Differentiation Engine \\

\#\#\#\# Task 3.1.1: Basic Differentiation Rules \\

- **Description:** Implement fundamental differentiation rules including power rule, product rule, quotient rule, and chain rule. Support differentiation of elementary functions and polynomial expressions. \\
- **Dependencies:** 2.2.1 \\
- **Difficulty:** 3/5 \\
- **Unit Tests:** \\
  - **Code Tests:** \\
    - **PowerRuleDifferentiation:** Test differentiation using power rule \\
    - **ProductRuleDifferentiation:** Test product rule implementation \\
    - **ChainRuleDifferentiation:** Test chain rule for composite functions \\

\#\#\#\# Task 3.1.2: Advanced Differentiation \\

- **Description:** Implement differentiation for transcendental functions, inverse functions, and implicit differentiation. Add support for partial derivatives and higher-order derivatives. \\
- **Dependencies:** 3.1.1 \\
- **Difficulty:** 4/5 \\
- **Unit Tests:** \\
  - **Code Tests:** \\
    - **TranscendentalDifferentiation:** Test differentiation of exponential and logarithmic functions \\
    - **TrigonometricDifferentiation:** Test differentiation of trigonometric functions \\
    - **PartialDerivatives:** Test partial differentiation for multivariate functions \\

\#\#\# Module 3.2: Integration Engine \\

\#\#\#\# Task 3.2.1: Basic Integration Methods \\

- **Description:** Implement fundamental integration techniques including power rule integration, substitution method, and integration by parts. Support integration of elementary functions and polynomial expressions. \\
- **Dependencies:** 3.1.1 \\
- **Difficulty:** 4/5 \\
- **Unit Tests:** \\
  - **Code Tests:** \\
    - **PowerRuleIntegration:** Test integration using power rule \\
    - **SubstitutionIntegration:** Test u-substitution method \\
    - **IntegrationByParts:** Test integration by parts \\

\#\#\#\# Task 3.2.2: Advanced Integration Techniques \\

- **Description:** Implement advanced integration methods including partial fractions, trigonometric substitution, and numerical integration fallbacks. Add support for definite integrals and improper integrals. \\
- **Dependencies:** 3.2.1 \\
- **Difficulty:** 5/5 \\
- **Unit Tests:** \\
  - **Code Tests:** \\
    - **PartialFractionIntegration:** Test partial fraction decomposition and integration \\
    - **TrigonometricSubstitution:** Test trigonometric substitution methods \\
    - **DefiniteIntegralEvaluation:** Test evaluation of definite integrals \\

\#\# Phase 4: Matrix Operations and Linear Algebra \\

**Goal:** Implement comprehensive matrix operations and linear algebra capabilities for solving systems of equations and performing matrix computations. \\

\#\#\# Module 4.1: Matrix Framework \\

\#\#\#\# Task 4.1.1: Matrix Data Structure \\

- **Description:** Implement a matrix class supporting symbolic entries, basic matrix operations, and efficient storage. Include support for sparse matrices and matrix metadata. \\
- **Dependencies:** 1.1.2 \\
- **Difficulty:** 3/5 \\
- **Unit Tests:** \\
  - **Code Tests:** \\
    - **MatrixCreation:** Test creation of matrices with various dimensions \\
    - **MatrixAccess:** Test element access and modification \\
    - **MatrixEquality:** Test matrix equality and comparison operations \\

\#\#\#\# Task 4.1.2: Basic Matrix Operations \\

- **Description:** Implement matrix addition, subtraction, multiplication, and scalar operations. Include support for matrix transposition and basic matrix properties. \\
- **Dependencies:** 4.1.1 \\
- **Difficulty:** 3/5 \\
- **Unit Tests:** \\
  - **Code Tests:** \\
    - **MatrixArithmetic:** Test matrix addition, subtraction, multiplication \\
    - **MatrixTransposition:** Test matrix transpose operation \\
    - **ScalarOperations:** Test scalar multiplication and division \\

\#\#\# Module 4.2: Linear Algebra Operations \\

\#\#\#\# Task 4.2.1: Matrix Decomposition \\

- **Description:** Implement LU decomposition, QR decomposition, and eigenvalue/eigenvector computation. Support symbolic computation where possible with numerical fallbacks. \\
- **Dependencies:** 4.1.2 \\
- **Difficulty:** 4/5 \\
- **Unit Tests:** \\
  - **Code Tests:** \\
    - **LUDecomposition:** Test LU decomposition for various matrix types \\
    - **QRDecomposition:** Test QR decomposition implementation \\
    - **EigenvalueComputation:** Test eigenvalue and eigenvector calculation \\

\#\#\#\# Task 4.2.2: System Solving \\

- **Description:** Implement Gaussian elimination, matrix inversion, and general linear system solving for systems with 10+ variables. Include support for under-determined and over-determined systems. \\
- **Dependencies:** 4.2.1 \\
- **Difficulty:** 4/5 \\
- **Unit Tests:** \\
  - **Code Tests:** \\
    - **GaussianElimination:** Test Gaussian elimination for various system sizes \\
    - **MatrixInversion:** Test matrix inversion for invertible matrices \\
    - **LargeSystemSolving:** Test solving systems with 10+ variables \\

\#\# Phase 5: Equation Solving and Numerical Methods \\

**Goal:** Implement comprehensive equation solving capabilities and numerical methods for root finding and approximation. \\

\#\#\# Module 5.1: Equation Solving Engine \\

\#\#\#\# Task 5.1.1: Polynomial Equation Solving \\

- **Description:** Implement solving algorithms for polynomial equations including quadratic formula, cubic and quartic formulas, and numerical methods for higher-degree polynomials. \\
- **Dependencies:** 2.1.2 \\
- **Difficulty:** 4/5 \\
- **Unit Tests:** \\
  - **Code Tests:** \\
    - **QuadraticSolving:** Test quadratic equation solving \\
    - **CubicSolving:** Test cubic equation solving using Cardano's formula \\
    - **HighDegreeSolving:** Test numerical solving of high-degree polynomials \\

\#\#\#\# Task 5.1.2: General Equation Solving \\

- **Description:** Implement solving algorithms for transcendental equations, systems of nonlinear equations, and symbolic equation manipulation. Include support for parametric solutions. \\
- **Dependencies:** 5.1.1, 3.1.2 \\
- **Difficulty:** 5/5 \\
- **Unit Tests:** \\
  - **Code Tests:** \\
    - **TranscendentalSolving:** Test solving of exponential and logarithmic equations \\
    - **NonlinearSystemSolving:** Test solving of nonlinear equation systems \\
    - **SymbolicSolving:** Test symbolic manipulation for equation solving \\

\#\#\# Module 5.2: Numerical Methods \\

\#\#\#\# Task 5.2.1: Root Finding Algorithms \\

- **Description:** Implement numerical root finding methods including Newton-Raphson, bisection method, and secant method. Support multi-dimensional root finding for systems of equations. \\
- **Dependencies:** 3.1.2 \\
- **Difficulty:** 3/5 \\
- **Unit Tests:** \\
  - **Code Tests:** \\
    - **NewtonRaphsonMethod:** Test Newton-Raphson root finding \\
    - **BisectionMethod:** Test bisection method implementation \\
    - **MultidimensionalRootFinding:** Test root finding for systems of equations \\

\#\#\#\# Task 5.2.2: Step-by-Step Solution Generator \\

- **Description:** Implement a system to generate detailed step-by-step solutions for all mathematical operations, including intermediate steps, explanations, and solution verification. \\
- **Dependencies:** 5.1.2, 3.2.2, 4.2.2 \\
- **Difficulty:** 4/5 \\
- **Unit Tests:** \\
  - **Code Tests:** \\
    - **StepGeneration:** Test generation of solution steps for various problem types \\
    - **SolutionVerification:** Test verification of generated solutions \\
    - **StepFormatting:** Test formatting of step-by-step solutions for display \\
\end{tcolorbox}

\begin{tcolorbox}[breakable, title=Task Decomposition for Project ``filesystem-simulator'': Task 1.1.1]
\small
\#\# Test Case 1: ExpressionNodeCreation \\

**Purpose:**  \\
Verify creation and initialization of all supported node types (variable, constant, operator, function) within the expression tree, including correct assignment of metadata and type information. \\

\#\#\# Test Scenarios \\

1. **Happy Path Cases:** \\
   - Create a variable node (e.g., 'x'). \\
   - Create a constant node (e.g., 4.2). \\
   - Create an operator node (e.g., '+', with appropriate left/right children). \\
   - Create a function node (e.g., 'sin', with argument node(s)). \\
   - Create a node with metadata (e.g., position, LaTeX source, type info). \\

2. **Edge Cases / Boundaries:** \\
   - Create a variable node with an empty string or invalid variable name. \\
   - Create a constant node with value zero, negative or extreme values. \\
   - Create an operator node with missing or null children. \\
   - Create a function node for a function with no arguments or multiple arguments. \\

3. **Error Handling:** \\
   - Attempt to create a node with an unsupported or invalid node type. \\
   - Attempt to create an operator node with an unknown operator symbol. \\
   - Pass null values to required fields and verify proper exception or error behavior. \\

\#\#\# Assertions \\

- Nodes are created with correct type and internal state. \\
- Metadata and type information are correctly assigned and retrievable. \\
- Operator and function nodes have correct child/argument associations. \\
- Proper exceptions or error messages are raised on invalid input. \\

--- \\

\#\# Test Case 2: ExpressionTreeTraversal \\

**Purpose:**  \\
Verify correct traversal of the expression tree using preorder, inorder, and postorder methods, including traversal order and node visitation. \\

\#\#\# Test Scenarios \\

1. **Happy Path Cases:** \\
   - Traverse a simple arithmetic expression tree (e.g., 'x + 2'). \\
   - Traverse a nested expression (e.g., 'sin(x\textasciicircum 2 + 1)'). \\
   - Traverse a tree with mixed operators and function nodes. \\

2. **Edge Cases / Boundaries:** \\
   - Traverse a single-node tree (e.g., just a constant or variable). \\
   - Traverse a highly unbalanced tree (e.g., only left or only right children). \\
   - Traverse a tree with unary operators or functions. \\

3. **Error Handling:** \\
   - Attempt traversal on an empty/null tree (should handle gracefully, no exceptions). \\
   - Traverse a malformed tree (e.g., operator node missing one child). \\

\#\#\# Assertions \\

- Traversal methods visit nodes in the correct order for each traversal type. \\
- All nodes are visited exactly once per traversal. \\
- Traversal outputs match expected node visitation sequences. \\
- Traversal methods handle empty and malformed trees without crashing. \\

--- \\

\#\# Test Case 3: ExpressionEquality \\

**Purpose:**  \\
Verify that two expression trees representing the same mathematical expression are considered equal, and that hashing is consistent and correct. \\

\#\#\# Test Scenarios \\

1. **Happy Path Cases:** \\
   - Compare two identical trees (structure and values). \\
   - Compare two trees with equivalent structure but different variable names or constant values (should not be equal). \\
   - Compare trees with different structures but representing the same expression (if structural equality is required, should not be equal; if mathematical equivalence, specify behavior). \\

2. **Edge Cases / Boundaries:** \\
   - Compare a tree to itself (reflexivity). \\
   - Compare a tree to an empty/null tree. \\
   - Compare deeply nested trees for correctness and performance. \\

3. **Error Handling:** \\
   - Attempt to compare a valid tree with an unsupported object type (should return false or raise TypeError). \\
   - Ensure hashing works for trees used as dictionary keys or set members. \\

\#\#\# Assertions \\

- Trees with identical structure and node values/types are equal. \\
- Trees with different structures or node values are not equal. \\
- Equality is reflexive, symmetric, and (where applicable) transitive. \\
- Hashes for equal trees are identical; hashes for non-equal trees differ. \\
- Comparing to null/invalid types does not raise unexpected exceptions. \\

--- \\

**Note:** These tests are distinct from planned future tests (e.g., no overlap with substitution/validation/copying or LaTeX parsing) and provide a robust foundation for the core expression data structure, as required for all dependent modules. \\
\end{tcolorbox}

\begin{tcolorbox}[breakable, title=Task Decomposition for Project ``filesystem-simulator'': Task 2.2.1]
\small
\#\# Test Case 1: ConstantFolding \\

**Purpose:**  \\
Verify that the simplification engine collapses constant expressions by evaluating arithmetic operations involving only constants, at all levels of the expression tree. \\

\#\#\# Test Scenarios \\

**Normal/Happy Path:** \\
- Simplify an expression with only constants: \texttt{2 + 3 * 4} $\rightarrow$ \texttt{14} \\
- Simplify constants in nested sub-expressions: \texttt{2 + (3 * (4 - 1))} $\rightarrow$ \texttt{11} \\
- Simplify powers and division with constants: \texttt{2\textasciicircum 3 / 4} $\rightarrow$ \texttt{2} \\

**Edge Cases / Boundaries:** \\
- Simplify a single constant (e.g., \texttt{5} $\rightarrow$ \texttt{5}) \\
- Simplify an expression where all terms are constants but involve different operators: \texttt{((2+2) * (3-1)) / 2} $\rightarrow$ \texttt{4} \\
- Simplify an expression with redundant parentheses: \texttt{((2))} \\

**Error Handling / Exception Cases:** \\
- Simplify an expression with division by zero: \texttt{6 / 0} (should raise or signal error in simplification step) \\
- Simplify an expression with invalid constant (e.g., \texttt{nan + 1}, if supported) \\

\#\#\# Assertions \\

- Result is a single constant node with the correct value for all-constant expressions. \\
- For nested expressions, constants are folded as deep as possible, and non-constant sub-expressions remain untouched. \\
- Division by zero or invalid arithmetic expressions raise a clear, descriptive exception or error flag. \\
- The simplification process does not mutate the original tree structure outside of constant folding. \\

--- \\

\#\# Test Case 2: IdentitySimplification \\

**Purpose:**  \\
Verify that simplification correctly applies algebraic identities to remove or reduce redundant operations (e.g., \texttt{x + 0} $\rightarrow$ \texttt{x}, \texttt{x * 1} $\rightarrow$ \texttt{x}, \texttt{x - 0} $\rightarrow$ \texttt{x}, \texttt{x / 1} $\rightarrow$ \texttt{x}, etc.). \\

\#\#\# Test Scenarios \\

**Normal/Happy Path:** \\
- Simplify addition and subtraction identities: \texttt{x + 0}, \texttt{0 + y}, \texttt{z - 0} \\
- Simplify multiplication and division identities: \texttt{x * 1}, \texttt{1 * y}, \texttt{z / 1} \\
- Simplify power identities: \texttt{x\textasciicircum 1}, \texttt{x\textasciicircum 0}, \texttt{1\textasciicircum y} \\
- Simplify nested identities: \texttt{(x + 0) * 1}, \texttt{(y - 0) / 1} \\

**Edge Cases / Boundaries:** \\
- Simplify expressions with multiple nested identities: \texttt{((x + 0) * 1) + 0} \\
- Simplify identity operations on constants: \texttt{0 + 0}, \texttt{1 * 1}, \texttt{0\textasciicircum 0} (should behave as specified by system conventions) \\
- Identity operations on function outputs (e.g., \texttt{sin(x) + 0}, \texttt{log(y) * 1}) \\

**Error Handling / Exception Cases:** \\
- Simplify invalid power identities such as \texttt{0\textasciicircum 0} (should raise exception or handle per convention) \\
- Division by one or subtraction of zero when sub-expressions are malformed (should not break simplification) \\

\#\#\# Assertions \\

- Simplified expressions match mathematically equivalent, minimal forms with redundant identities removed. \\
- Complex expressions simplify recursively (identities are collapsed wherever they appear). \\
- Special cases like \texttt{x\textasciicircum 0} yield \texttt{1}, \texttt{0\textasciicircum x} yields \texttt{0} for \texttt{x} $\neq$ \texttt{0}, etc., handled according to system conventions. \\
- Edge cases such as \texttt{0\textasciicircum 0} are handled as per system design (error or defined value). \\
- Original expressions are not mutated unless simplification is performed in-place by design. \\

--- \\

\#\# Test Case 3: ZeroOneSimplification \\

**Purpose:**  \\
Test simplification rules that specifically target algebraic behaviors involving zero and one, such as multiplication and division by zero/one, and additive/multiplicative annihilation. \\

\#\#\# Test Scenarios \\

**Normal/Happy Path:** \\
- Simplify multiplication by zero: \texttt{x * 0}, \texttt{0 * y}, \texttt{(a + b) * 0} $\rightarrow$ \texttt{0} \\
- Simplify division of zero: \texttt{0 / x} $\rightarrow$ \texttt{0} for \texttt{x} $\neq$ \texttt{0} \\
- Simplify division by zero: \texttt{a / 0} (should raise or signal error) \\
- Simplify expressions with multiple zeros or ones: \texttt{0 + 0}, \texttt{1 * 1}, \texttt{(x * 1) + 0} \\

**Edge Cases / Boundaries:** \\
- Simplify zeroes as function arguments or return values: \texttt{sin(0)}, \texttt{log(1)}, etc. (if included in basic simplification) \\
- Simplify expressions where zero or one is nested inside functions or deep sub-expressions. \\
- Simplify multiplication or division chains: \texttt{x * 0 * y}, \texttt{(0 / x) / y}, \texttt{(x / 1) * 1} \\

**Error Handling / Exception Cases:** \\
- Division by zero (\texttt{a / 0}, \texttt{0 / 0}) is detected and handled per system convention (error, undefined, etc.). \\
- Chained or redundant zero/one operations are fully simplified (e.g., \texttt{((x * 1) + 0) * 0} $\rightarrow$ \texttt{0}) \\
- Propagation of errors through the simplification (e.g., if one sub-expression involves division by zero, the whole expression should reflect the error state) \\

\#\#\# Assertions \\

- All cases of multiplication by zero result in \texttt{0}. \\
- Division of zero by non-zero yields \texttt{0}. \\
- Division by zero is detected, flagged, or raises an exception as specified. \\
- Chains of zero/one produce the correct simplified results recursively. \\
- Zero and one simplification rules do not interfere with other simplification rules or mutate unrelated subtrees. \\

--- \\

**Note:** \\
- These tests strictly address the simplification and rewriting logic, not arithmetic or structural manipulation already covered by previous tests. \\
- They depend on the ability to traverse and rewrite expression trees, leveraging previously-tested traversal and node creation. \\
- Results should be checked for mathematical correctness and structural integrity (i.e., tree correctness is preserved after simplification). \\
- Error and exception propagation during simplification must be robust and informative. \\
\end{tcolorbox}

\end{document}